# Auroral imaging with combined Suomi 100 nanosatellite and ground-based observations: A case study

Esa Kallio[1], Ari-Matti Harri[2], Olli Knuuttila[1], Riku Jarvinen[1,2], Kirsti Kauristie[2], Antti Kestilä[2], Jarmo Kivekäs[2], Petri Koskimaa[2], Juha-Matti Lukkari[1], Noora Partamies[3], Jouni Rynö[2], and Mikko Syrjäsuo[3]

[1] Aalto University, School of Electrical Engineering, Espoo, Finland

[2] Finnish Meteorological Institute, Helsinki, Finland

[3] The University Centre in Svalbard, Birkeland Centre for Space Science, University of Bergen, Norway

Corresponding author: Esa Kallio (esa.kallio@aalto.fi)

**Key Points:**

- The concept of imaging aurora towards the Earth's limb by a cubesat camera is demonstrated.

- The dark background available in the Earth-limb viewing direction facilitates imaging of dim auroras e.g., for auroral tomography purposes.

- The analysis shows that auroral imaging at low Earth orbit can provide a new context for ground-based auroral and ionospheric observations.

## Abstract

Auroras can be regarded as the most fascinating manifestation of space weather and they are continuously observed by ground-based and, nowadays more and more, also by space-based measurements. Investigations of auroras and geospace comprise the main research goals of the Suomi 100 nanosatellite, the first Finnish space research satellite, which has been measuring the Earth's ionosphere since its launch on Dec. 3, 2018. In this work, we present a case study where the satellite's camera observations of an aurora over Northern Europe are combined with ground-based observations of the same event. The analyzed image is, to the authors' best knowledge, the first auroral image ever taken by a cubesat. Our data analysis shows that a satellite vantage point provides complementary, novel information of such phenomena. The 3D auroral location reconstruction of the analyzed auroral event demonstrates how information from a 2D image can be used to provide location information of auroras under study. The location modelling also suggests that the Earth's limb direction, which was the case in the analyzed image, is an ideal direction to observe faint auroras. Although imaging on a small satellite has some large disadvantages compared with ground-based imaging (the camera cannot be repaired, a fast moving spinning satellite), the data analysis and modelling demonstrate how even a small 1-Unit (size: 10 cm × 10 cm × 10 cm) CubeSat and its camera, build using cheap commercial off-the-shelf components, can open new possibilities for auroral research, especially, when its measurements are combined with ground-based observations.

## 1 Introduction

Auroras form when high-energy charged particles from the magnetosphere precipitate and collide with atmospheric neutrals at about 100 – 200 km altitude. Therefore, auroras are manifestations of energization processes occurring in the near-Earth space. Auroras are investigated from the ground by many different types of instruments. Among the most common instruments used to investigate auroras are ground-based all-sky cameras (ASCs) which can have a large Field-of-View (FoV), by default 180°, therefore, offering a good probability to capture auroras (e.g., Mathews et al., 2004). However, the effective usage of such optical data is reduced by cloud coverage, which means that an interesting auroral emission may not be effectively imaged, even though the camera devices are in operation.

A camera onboard a satellite which flies above cloud layers, however, can capture auroras under any terrestrial weather conditions. Moreover, the auroral light imaged from above in space does not propagate through the low-altitude, high-density atmosphere as it does when it propagates to an ASC on the ground. Therefore, the auroral light is less attenuated by the atmosphere (Oikarinen, 2001), when measured from space rather than being viewed from the ground.

The total number of auroral cameras in space is, however, much lower than the number of ground-based auroral cameras due to the high costs of space-based instruments and satellite missions. Especially, a camera instrument on a satellite can very seldom be repaired and, therefore, the risk of the permanent loss of the instrument makes it less likely that expensive high-performance cameras will be added to spacecraft. Ground-based optical-instrument networks are typically built to support satellite observations of aurora-related processes, as in the case of the THEMIS (Mende et al., 2009) and ARASE missions (Shiokawa et al., 2017), or to provide a wider context for radar observations, as in the case of EISCAT_3D (Ogawa et al. 2020). Stable ground-based instruments (e.g. Syrjäsuo, 2001; Brändström, 2003; Sangalli et al., 2011) can provide continuous systematic observations reliably for many years from the same location on the ground allowing long-term statistical studies to be conducted (see e.g., Shiokawa et al., 2019; Partamies et al., 2014). A camera instrument mounted on a rocket provides another possibility to measure auroras in different wavelengths (see e.g., Ellingsen et al., 2015). The time during which rocket observations are available is, however, very limited.

A nanosatellite, i.e., a satellite in the mass range 1 – 10 kg, instead, provides a potential to increase the number of scientific cameras in space because of their relatively small construction, launch and operation costs compared to more traditional satellites with typical masses well over 100 kg. An example of one of the first small 1-Unit CubeSat (10 cm × 10 cm × 10 cm) satellites with a camera designed to investigate the Earth's atmosphere, its airglow, was the SwissCube-1 satellite. The instrument took photographs in a narrow 762 nm channel from March to December 2011 (Rossi et al., 2013). Increasing the size of the nanosatellite, in turn, enables a more dedicated space research investigation, such as 3-Unit Miniature X-Ray Solar Spectrometer (MinXSS) CubeSats which were designed to perform Solar soft X-Ray measurements (Moore et al., 2018). Most recently, a 2-Unit AMICal has taken its first auroral image (Barthelemy et al., 2022) while another dedicated auroral imaging satellite, a 12-Unit ATISE (Barthelemy et al., 2018) is awaiting launch.

A small satellite has natural limitations due to its mass, volume, power, telecommunication rate as well as, typically, camera pointing accuracy that is relatively limited. Therefore, it is not clear, without conducting proper analysis of the obtained data, what the benefits and drawbacks of a camera instrument on a nanosatellite actually are compared to a camera on a large satellite, or a camera on the ground.

In this work we present a case study where auroras were photographed on January 22[nd], 2019, with the Suomi 100 satellite, the first Finnish research satellite, and these observations were correlated with ground-based camera and magnetic field measurements. We study how the satellite's measurements can be analyzed and interpreted despite the specific limitations of nanosatellites.

The paper is organized as follows. First, we introduce the Suomi 100 satellite and its white-light wide-angle camera instrument as well as the used ground-based instruments. Second, we introduce the image taken by the satellite while pointing towards the Norwegian sea. Third, we combine the satellite's auroral image with the ground-based observations. Fourth, we analyze the observation with a FoV model which is used to reconstruct the possible positions of the observed auroras. Finally, the results and lessons learned are presented.

## 2 Instruments

### 2.1 Suomi 100 satellite and its camera

The Suomi 100 satellite was launched on December 3[rd], 2018, to a polar Sun Synchronous orbit (SSO) at about 600 km altitude. The satellite is a 1-Unit CubeSat with the size of 10 cm × 10 cm × 10 cm and a mass of 1.3 kg. The research goal of the satellite is to investigate auroras and the ionosphere with a payload consisting of two instruments, a camera and a radio spectrometer. The camera is a 3-megapixel (2048×1536 pixels) RGB (Red, Green and Blue) wide-angle camera with a focal length of 8.2 mm and a pixel pitch of 3.2 μm giving a horizontal×vertical FoV of 43.6°×33.4° and a ground resolution of about 235 m (see Knuuttila et al., 2022, for the details of the camera instrument). The radio spectrometer was designed to investigate ionospheric electron density both by making independent measurements and by making joint observations with the European Incoherent Scatter Scientific Association (EISCAT) High Frequency (HF) facility (Kallio et al., 2022).

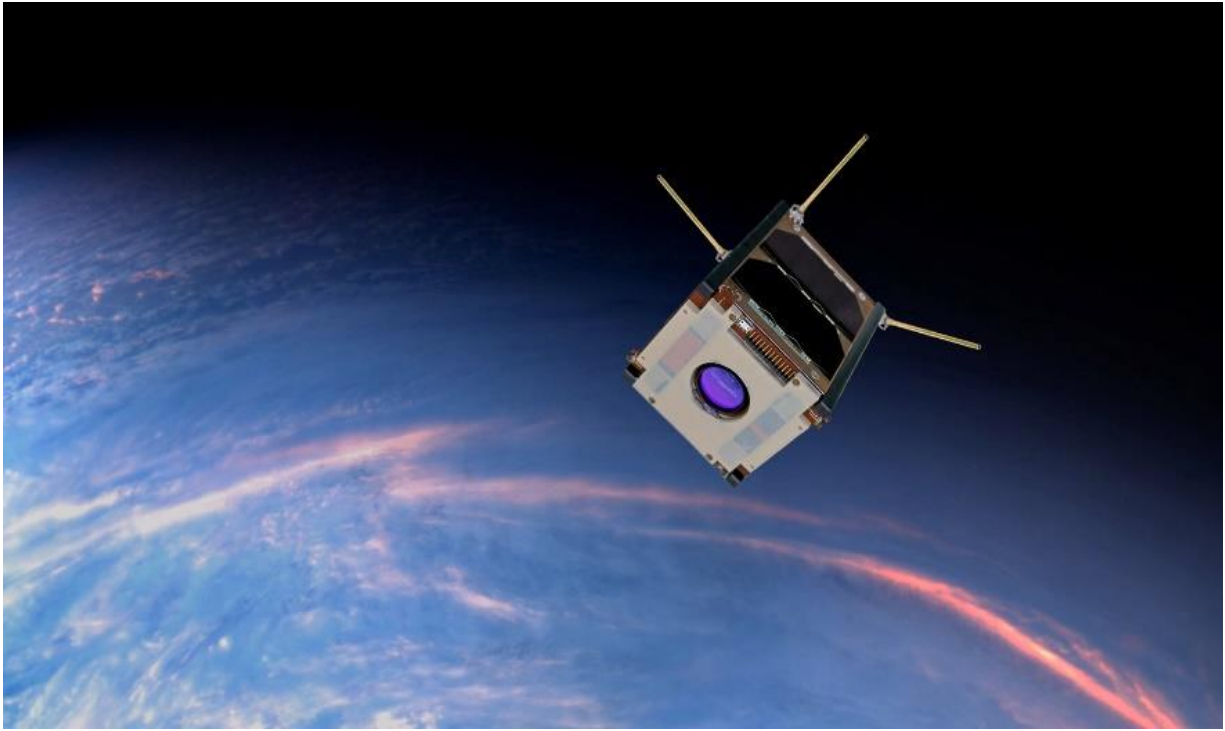

**Figure 1**. A photomontage of the Suomi 100 satellite photographed on the ground and a part of a combination image made of two Earth photos (#687, #688) taken by the Suomi 100 satellite on December 9th, 2018, soon after the launch. The objective of the camera can be seen at the center of the bottom light color face of the satellite, which typically points near the nadir. The orange color band on the bottom right is associated with the light scattering from the tropopause jet stream.

The camera was operated in two phases. In the first phase, the camera was used on the Earth's dayside in order to initially validate that the camera was intact after the stresses caused by the launch of the spacecraft. Analysis of sharp daylight images confirmed that the focus of the objective was unaffected by the launch. Further, radiometric calibration of the camera was conducted in detail in space (Knuuttila et al., 2022). The images also showed that the JPEG (Joint Photographic Experts Group) format, i.e., the data format which has been so far used to transfer images to ground, worked as in the ground calibrations and did not cause artefacts. The camera can also take non-compressed raw images, which have not been used so far because of the low telemetry rate. When an image is taken, the satellite automatically saves a small 128×96 pixel version of it, called a thumbnail. During satellite operations, thumbnails are first downloaded and then, if any appear to be of scientific interest by visual inspection, such as including celestial objects, which can be used for a camera calibration (see Knuuttila et al., 2022) or, especially, if they possibly include auroras, the corresponding larger JPEG image will be subsequently downloaded.

The camera could not be accurately directed at any given time because of the satellite's relatively simple attitude control system, which consisted of magnetically actuated de-tumbling. Instead, many images were taken in order to maximize the probability that at least some of them would be taken while pointing in a desirable direction. The slow rotation of the satellite which, for example, in the analyzed auroral image case was about one rotation per 2 minutes, made it possible to make a panoramic image ranging from the limb of the Earth to the day-night terminator region (Figure 2).

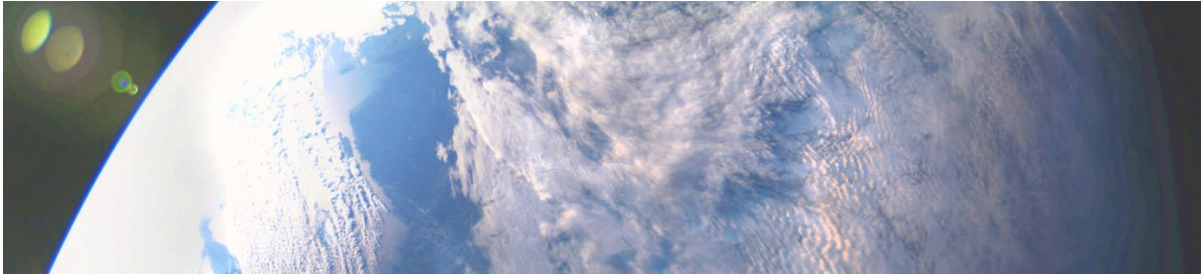

**Figure 2**. An example of the dayside images used to confirm the quality of the Suomi 100 satellite's imaging pipeline. This panoramic image is a photomontage combined from 5 images (#1092 - #1096) taken on January 3[rd], 2019, each one taken when pointing in slightly different directions during the rotation of the spacecraft. The brightest images are balanced, and the features of the figures are slightly deformed in order to combine the images smoothly together. The figure shows a rectangular part of the full panorama image. Note that the black region on the left shows the space above the local horizon at the position of the spacecraft at about 600 km altitude, while the dark region on the right-hand side of the figure shows the Earth's nightside.

After the dayside imaging phase, a second, nightside imaging phase was started after adjusting the camera settings, such as exposure time and gain, to allow the imaging of auroras. Similarly, to the dayside image analysis, thumbnails were first downloaded and if they appeared interesting, larger sized JPEG images were subsequently downloaded.

### 2.2 Ground-based instruments

In this paper, Suomi 100's observations are combined together with observations from three types of ground-based instruments: [1] the Longyearbyen meridian scanning photometer (MSP) at Kjell Henriksen Observatory (KHO), Norway; [2] an all-sky DSLR at KHO on Svalbard, Norway; and [3] magnetometers on Svalbard, Norway. These observations were chosen because of their close proximity to the position of the imaged auroras.

The full color ASC images from KHO in Svalbard (78.20°N, 15.82°E) were captured by a Sony α7S camera with fish-eye optics. The images were taken with a 4-second exposure and a cadence of 11–12 seconds. Each image size is 2832×2832 pixels, and the images are oriented so that the geomagnetic north is at the top of the image and east is to the left of the image.

The MSP under the same roof provides auroral spectral intensities at 630.0, 427.8, 557.7 and 844.6 nm, which correspond to atomic oxygen emissions and an emission from molecular nitrogen ions (427.8 nm). Each channel is equipped with an interference filter with a bandpass of 0.5 nm (FWHM). Using a rotating mirror, this data is collected along a geomagnetic meridian scan with a 1-degree FoV. The temporal resolution of the data is 16 seconds (a detailed description of the two optical instruments and their measurements can be found in Dreyer et al., 2021).

The triaxial fluxgate magnetometers in the Svalbard area are owned and operated by Tromsø Geophysical Observatory (TGO) at UiT, The Arctic University of Norway. The stations located at Ny-Ålesund (78.92°N), Longyearbyen (78.20°N), Hopen (76.51°N) and Bjørnøya (74.50°N) are also part of the International Monitor for Auroral Geomagnetic Effects (IMAGE) magnetometer network (Tanskanen, 2009). These measurements have a 10-second temporal resolution, and in this study, we show the horizontal component of the magnetic field deflections, which is indicative of enhancements in horizontal ionospheric currents and is caused by auroral particle precipitation.

## 3 Observations

### 3.1 Suomi 100 satellite

The clearest photograph of the aurora taken by the Suomi 100 satellite so far comes from January 22$^{nd}$, 2019, at about 18:03:13 UT. In this article a cleaned version of the image is used, where the on-board gamma and color corrections have been inverted and the warm pixel induced structural noise removed by subtracting an image taken the next day of the Orion constellation that approximates a dark image (see Knuuttila et al., 2022, for details). This processed image is shown in the bottom panel of Fig. 3. and it is referred to in this paper as the auroral image, the auroral event or simply as the image. Additionally, Fig. 3 shows the orbit of the satellite about 4 minutes before and after taking the image, at approximately 62.17°N 47.64°E altitude 582.9 km. Moreover, the auroral image was taken relatively close to the winter solstice at a local time around 9 pm (18 UT), which provides favorably dark conditions (solar elevation -38.1°). At the time of the auroral image, the satellite's rotation period was about 2 minutes. Therefore, the region of the aurora captured in the auroral image at 18:08:13 UT, is nearly at the center of the image, while one rotation period earlier (at 18:06:02 UT), it was more to the right of the image. Note that because the velocity of the satellite was about 7 km/s the satellite travels approximately 840 km in 2 minutes.

The auroral image has some notable features associated with it, not only the measurement technique, but also the measurement conditions and the auroral physics. First, the exposure time of the auroral image was 2.4 seconds while the rotation time of the spacecraft was about 2 minutes. Consequently, the satellite rotated an angle of about 7.2° (= (2.4 s / 120 s) × 360°) during the exposure. Therefore, the image contains some motion blur: for instance, the rotation of the light spots on the ground in the lower part of the image, especially near Murmansk. The rotation of stars are also visible, especially the bright 0.76 magnitude Altair (the white colored star) and 2.7 magnitude Tarazed (the orange-hued colored star) above it in the Aquila constellation. Second, the lower part of the auroral image is relatively bright (although it is night because of the light scattering from the clouds over Northern Scandinavia and Russia). Third, the aurora can be identified near the center of the image above the horizon as a vertically-extending relatively narrow emission pattern caused by electron precipitation along the magnetic field. The non-structured colored horizontal layers above the horizon are, instead, nightglow.

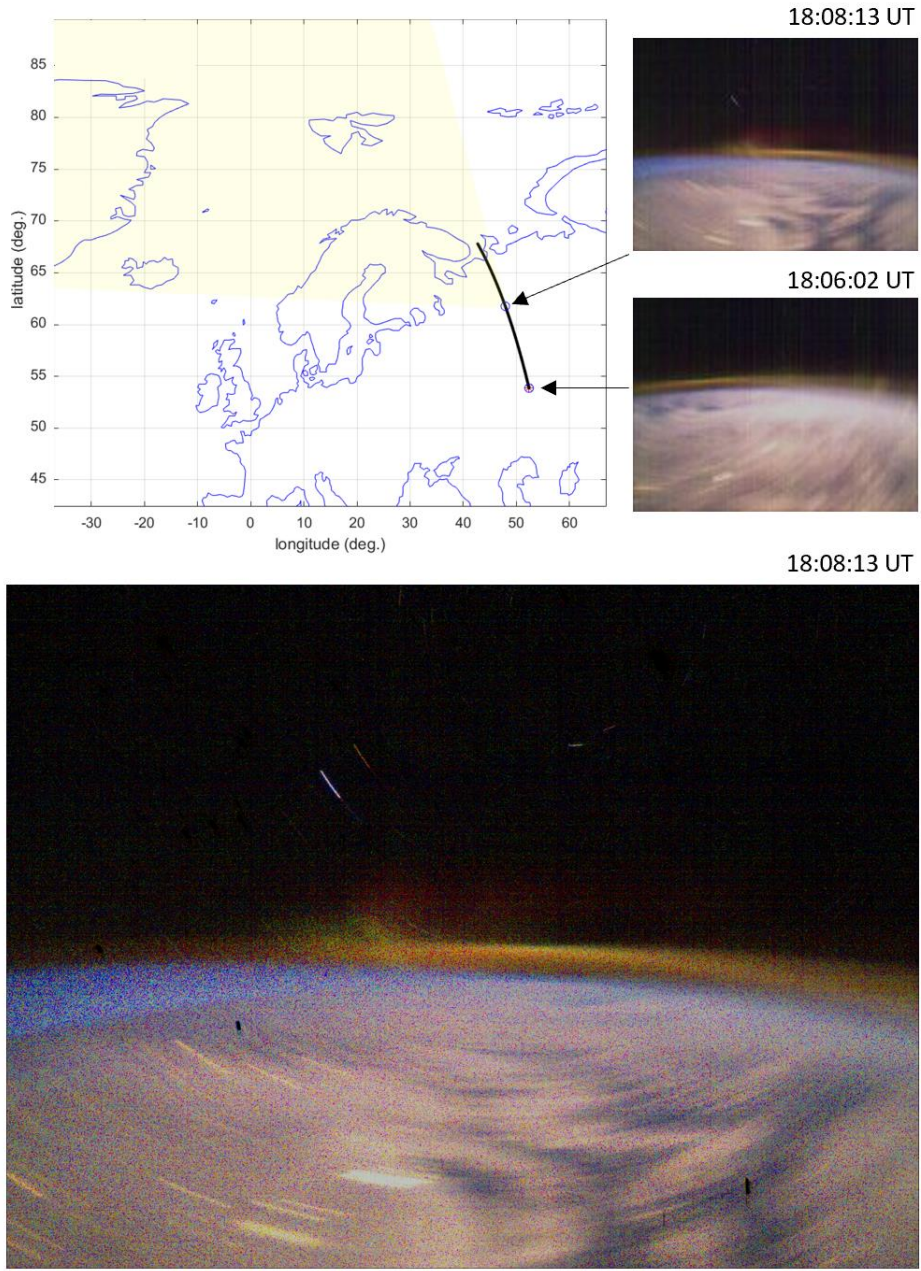

**Figure 3**. Images of the aurora taken by the Suomi 100 satellite on January 22$^{nd}$, 2019. Top left: The path of the satellite between 18:06:02 UT – 18:09:55 UT in a latitude-longitude map. The circles on the orbit show the position of the satellite at 10-minute time intervals. The slight yellow shading shows an approximate field-of-view of the image taken at 18:08:13 UT, as will be discussed later in Section 4. Two small inserts on the top right: A thumbnail (18:06:02 UT, #1224_thumb) of an image taken about two minutes before the analyzed auroral image, whose thumbnail is on the top right (18:08:13 UT, #1229_thumb). The satellite rotates so that the horizon in the images rotates clockwise. Bottom: The full-size auroral image (18:08:13 UT, #1229) where the aurora is located near the center of the image above the horizon.

### 3.2 Ground-based observations

Fig. 4b-d show the ground-based observations from Svalbard, Norway, and Sodankylä, Finland, near the time when the auroral image (Fig. 4a) was taken.

In the Longyearbyen/KHO MSP observations (Fig. 4b), an intense emission in the red (630.0 nm) and green (557.7 nm) channels is seen over the southern part of the sky at about 00:00 – 04:00 UT and about 18:00 – 24:00 UT, and in the north at other times on January 22nd, 2019. These two wavelengths map predominantly to the red and green channels of the satellite's camera, respectively. As one can see in Fig. 4a, and as will be analyzed in more detail in the model (Section 4), the direction of auroras in the auroral image is southward when viewed from Longyearbyen. At about 18:00 UT, Longyearbyen MSP observed an intense red and faint green light emerging from the south. Prior to that, a weak red aurora was observed to extend across the sky from north to south.

**Figure 4**. Ground-based observations from Svalbard and Sodankylä near the time of the auroral image. (a) The auroral image at 18:08 UT where an approximate geographic map is overlaid together with the positions of Longyearbyen and Tromsø. The white dotted line shows the approximate southward part of the scan of the Longyearbyen MSP (see Fig. 5 figure caption for details). (b) Meridian scanning photometer observations from Longyearbyen, Svalbard, Norway, at 17:30 – 18:30, on January 22nd, 2019. The panels show emissions at the wavelengths of 557.7 nm (green, top) and 630.0 nm (red, bottom), respectively. The plots show all scan angles from 0° (magnetic north) to 180° (magnetic south) and the color maps gives counts. Note that while MSP is not operated when the moon is directly in its field-of-view, any scattered light is removed from the data by background subtraction on every scan. (c) Auroral electrojet activity as derived from IMAGE ground-based magnetometer data with the method of Spherical Elementary Current Systems (SECS) method. The horizontal axis shows the event period: 18:06:00 – 18:09:00 UT on January 22nd, 2019, with a time resolution of 10 s. Note the slight intensification of the eastward currents (positive values) at the magnetic latitude around ~74° near the time when the auroral image was taken (18:08:13 UT). The quiet time period 18:05:00 – 18:05:50 UT is used for the baseline subtraction. The values are derived along the geographic longitude 22°E. (d) Three all-sky camera (ASC) images from KHO, Svalbard,

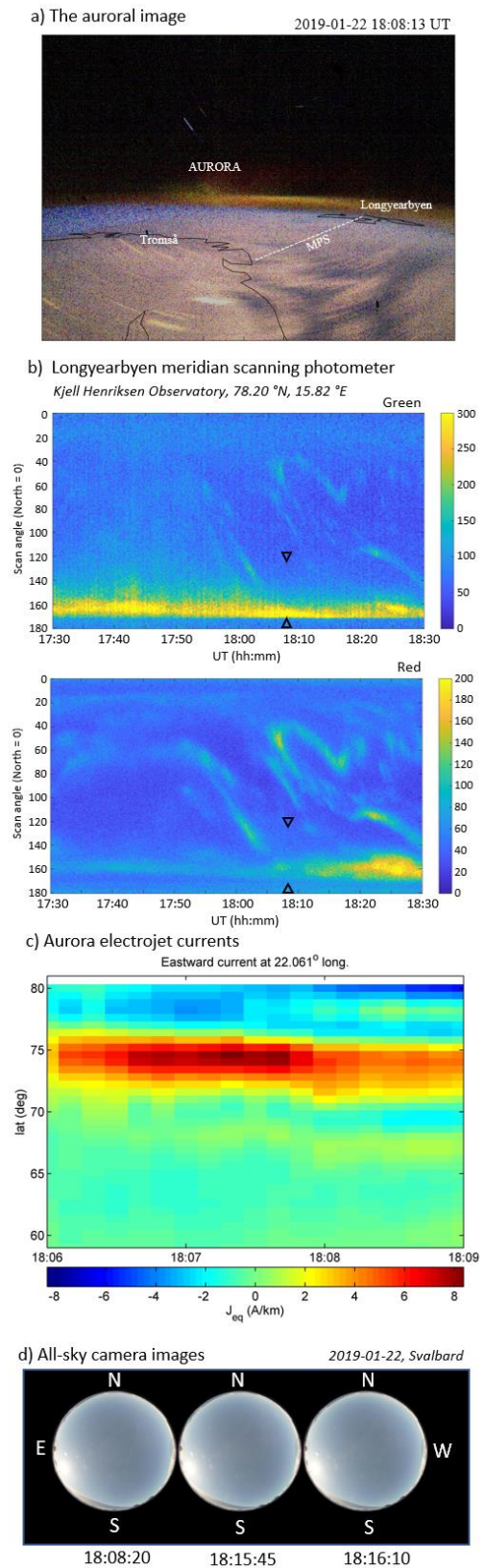

a) The auroral image      2019-01-22 18:08:13 UT

AURORA

Tromsø     MSP     Longyearbyen

b) Longyearbyen meridian scanning photometer
*Kjell Henriksen Observatory, 78.20 °N, 15.82 °E*

Green

Red

c) Aurora electrojet currents
Eastward current at 22.061° long.

$J_{eq}$ (A/km)

d) All-sky camera images      2019-01-22, Svalbard

18:08:20      18:15:45      18:16:10

on 2019-01-22 between 18:08:20 – 18:16:10 UT. The direction of the magnetic North, East, South, and West are marked by the letters N, E, S, and W, respectively. The exposure time was 4 s. As seen later in the analysis shown in Chapter 4, the auroras are expected to be seen on the southern part of the sky (in Fig. 4b in the South between the black arrows at scan angles 140° – 180°) and, correspondingly, in Fig. 4d low on southwestern sky.

The analyzed event was also investigated from the perspective of auroral electrojet currents. A weak intensification of eastward current is detected at 18:06 – 18:09 UT (Fig. 4c) in the estimation of ionospheric currents by the Spherical Elementary Current Systems (SECS) method (Amm and Viljanen, 1999) using the IMAGE magnetometer observations. The intensification is visible when the magnetic field recordings during 18:05:00 – 18:05:50 UT are considered as the baseline situation and the elementary current intensities after that time period are derived using magnetic field deviations from the given baseline.

Finally, also ground-based ASC images were investigated. The analyzed event is not ideal for ASC imaging on Svalbard because of the bright Moon in the south-eastern part of the sky (Fig. 4d). No clear auroras can be seen towards the south at the time of the auroral image.

In addition to the observations shown in Fig 4b-d, data from a longer time range and other ground-based instruments were investigated. First, no auroras are to be seen in the Svalbard's ASC images in Fig 4b. Green emission can be identified in the images in the south only at 19:35:10 UT (Supporting material S01). Second, multi-instrument analysis was further extended by utilizing the magnetometer observations from four stations of the Tromsø Geophysical observatory: Ny-Ålesund, Longyearbyen, Hopen Island, and Bear Island (Supporting material S02). A clear activation was detected after about 19 UT in the horizontal (H) component of the magnetic field at all the four stations. At two of the stations, Ny-Ålesund and Longyearbyen, the activation starts already at about 18:00 UT. The timing of these activations suggests that they can be associated with the auroral event captured by the Suomi 100 satellite.

Third, the magnetic field analysis was further elaborated by investigating the magnetic field XYZ components from the IMAGE magnetometer stations (Supporting material S03). Investigation of six magnetometers at the range of geographic latitudes ~ 70° – 77°, geographic longitudes ~ 15° – 25°, does not reveal any substantial activity around 18:00 UT, although some small activity was detected on the X, Y and Z components in one of the stations (Bjørnøya).

Forth, also ASC images from Kilpisjärvi (69.05°N) and Kevo (69.76°N), Finland, at around 18 UT were investigated, as well as an ASC keogram from Sodankylä (67.42°N), Finland. At Kilpisjärvi, there was some snow on the top of the ASC dome, but the zenith direction was clear. No obvious auroras were detected at 18:00 UT, but a later image at 18:59 UT contains a faint auroral arc (Supporting material S07). The situation at Kevo was quite similar: No auroras were detected at 18:00 UT but later, about 18:48 UT, an auroral arc, which brightened from east to west, was detected (Supporting material S06). In the ASC keogram from Sodankylä, a clear brightening was identified at around 19:00 UT and a weak increase of intensity before that, but no bright auroras at 18:00 UT (Supporting material S04).

Also several other observations were investigated. Sodankylä Geomagnetic Observatory's riometer observations from six stations were examined (Abisko: 30 MHz, 68.40°N 18.90°E; Ivalo: 30 MHz, 68.55°N 27.28°E; Sodankylä: 30 MHz, 67.42°N 26.39°E; Rovaniemi: 32.4 MHz, 66.78°N 25.94°E; Oulu: 30 MHz, 65.08°N 25.90°E; Jyväskylä: 32.4 MHz, 62.42°N 25.28°E)

(Supporting material S05). A riometer (Relative ionospheric opacity meter) monitors properties of the ionosphere by measuring how strongly a few tens of MHz radio waves are attenuated when they travel from space to the ground through the ionosphere. Variations in the riometer data can be associated, for example, with high energy particle precipitation. At Ivalo, fluctuation in the data existed before about 18:00 UT, but in all other stations no disturbances at that time were detected.

Finally, possible auroral images and magnetometer observations from Greenland were inspected. No ASC data to the west and north of Greenland were found. Furthermore, no magnetometer data was available from the East Greenland magnetometer chain at the time and latitude of the auroral image. The magnetometer data from the West Greenland chain was available, but it did not show any clear magnetic activity, suggesting that the auroral event took place between Svalbard and Northern Fennoscandia rather than further west over Greenland.

## 4 Modeling of observations

Interpretation of the Suomi 100 auroral image is complicated by the evolution of the satellite position with respect to the ground, the pointing direction of the camera, and the attitude of the camera.

In order to combine information from the auroral image and the ground-based observations, the position of the auroral event was investigated in this paper by developing a 3D auroral position model. In the model, one can mimic the positions of the auroral emission. A virtual camera is then employed to provide synthetic auroral intensity for images both in orthographic and perspective projection. In the model, the position of localized auroral emission spots can be placed at given altitudes with given longitude and latitude values. The virtual camera with a FoV mimicking the Suomi 100 satellite is then added with the desired position and orientation. The emission of red, green and blue light was simulated separately by using red, green and blue emission positions, respectively.

The positions of the auroral emission for the given analyzed wavelength, $\lambda$, photons was generated by using the photon emission density rate, $q_\lambda(\mathbf{r})$ [photons/m$^3$/s], where the altitude($h$) dependence was modeled by using the Chapman production function, $f_\lambda{}^{Chapman}$,

$$f_l^{Chapman}(h) = e^{(1 - y_l - e^{-y_l})}, \qquad (1a)$$

$$y_l = (h - h_l^{max})/H_l, \qquad (1b)$$

at a given latitude($\theta$) and longitude($f$)

$$q_l(\mathbf{r}) = g_l(\theta, f)\, f_l^{Chapman}(h), \qquad (2)$$

where $h_l^{max}$ and $H_l$ are the wavelength's peak emission altitude and scale height, respectively. In other words, the probability that a photon is formed in a volume dV during a time interval depends on the value $q_\lambda(\mathbf{r})$ dV. Note that the maximum value of the Chapman function in Eq. 1a is 1 at the peak altitude. Detailed determination of the altitude of the green and red auroras based on ground-based images is a challenging task (see e.g., Whiter et al., 2013). In this work, the following values were adopted for the red (630.0 nm), green (557.1 nm) and blue (427.8 nm) wavelength emissions: $h_{630.0\,\text{nm}}^{max} = 230$ km, $h_{557.1\,\text{nm}}^{max} = h_{427.8\,\text{nm}}^{max} = 110$ km, $H_{630.0\,\text{nm}} = 50$ km, $H_{557.1\,\text{nm}} = H_{427.8\,\text{nm}} = 25$ km. Note that for geometric analysis, we assume that the red channel activation above the Earth's limb is only due to the emission at 630.0 nm, even though $N_2^+$ is known to emit at wavelengths that would also activate the red channel. A similar assumption has given good results before [Partamies et al., 2007]. The latitude, longitude function was assumed to be a non-zero constant in the region where the emission was assumed to be non-zero so that the relative magnitudes were $g_{630.0\ nm} / g_{557.1\ nm} / g_{427.8\ nm} = 0.5/1/0.2$. The red to green maximum emission ratio $g_{630.0\ nm} / g_{557.1\ nm} = 0.5/1 = 1/2$ is used as an order of magnitude approximation for the estimated ratio (see Qi et al., 2017, Fig. 2, and Jackel et al., 2003, Fig. 5) similarly as the used green to blue maximum emission ratio $g_{557.1\ nm} / g_{427.8\ nm} = 1/0.2 = 5$ (see Enell et al., 2012, Fig. 6).

The total photon production rate, $Q_\lambda$ [photons/s], in the analyzed latitude-longitude region was finally obtained by integrating the photon density emission rate over the volume where the aurora was modelled to exist

$$Q_l = \int_{h_{min}}^{h_{max}} q_l(\boldsymbol{r}) = \int_{h_{min}}^{h_{max}} f_l^{Chapman}(h)\ r^2\ dh \iint g_l(\Theta, f)\ dW. \qquad (3)$$

Here the double integration with respect to the solid angle is over the analyzed latitude-longitude region. In this study, the emission altitude range, $[h_{min}, h_{max}]$, was assumed to be [50, 500] km.

In the model, a numerical technique was chosen where all three emission wavelengths were assumed to include the same amount of emission source points, $N_{souce}$. This technique makes it possible to obtain similar spatial statistics even if the total photon production rate of some wavelength would be much smaller than the total photon production rate of other wavelengths. Because all wavelengths have the same amount of emission points, the number of emitted photos from a source point per a unit time, $w_l$, depends on the analyzed wavelength

$$w_l = Q_l / N_{source}. \qquad (4)$$

In the used Chapman function the relative emission value $w_{630.0\ nm} / w_{557.1\ nm} / w_{427.8\ nm}$ was 1/0.96/0.19. This indicates that the total number of photons per source position emitted from red and green aurora is quite similar. The total number of photons per source position emitted from blue aurora is, instead, about 20 % of the photons emitted from the red and green positions.

Fig. 5 shows the reconstruction of the possible latitude and longitude positions at which the imaged aurora could have been. For simplicity, in this first reconstruction, the satellite's camera was assumed to have a circular FoV of 53°, mimicking the diagonal FoV of the Suomi 100 satellite's camera. The camera was assumed to look in the direction where auroras were detected in the auroral image (near the center of the image). The orientation of the camera was defined by using the position of the spacecraft and bright stars in the Aquila constellation, which was behind the auroral emission in the auroral image.

In the model, $10^7$ green and $10^7$ red emission locations were generated randomly around the whole Earth which modeled the emission of the red, green and blue auroras. Note that the spatial distribution of the green positions shows also how the spatial distribution of the blue emission positions would look like because they have an identical probability distribution in the model. Note also that Fig. 5 shows only a part of the collected emission points, $4 \times 10^4$ green points and $4 \times 10^4$ red points, so that both the geographic map and FoVs of the ground-based instruments would be visible.

The main difference in Fig. 5 between the locations of the green and red auroras is that the red auroras could have been observed almost anywhere above Greenland because red auroras originate from higher altitude than green auroras. It is noteworthy that the position of the spacecraft and the attitude of the camera were ideal to make joint auroral measurements with the ground-based instruments in Fennoscandia and towards the North Pole (because the camera could have seen the same auroras as observed by ASCs at Longyearbyen, Kevo and Kilpisjärvi as well as Longyearbyen MSP). Moreover, the camera looked towards north which, in January, provided a good permanent dark sky.

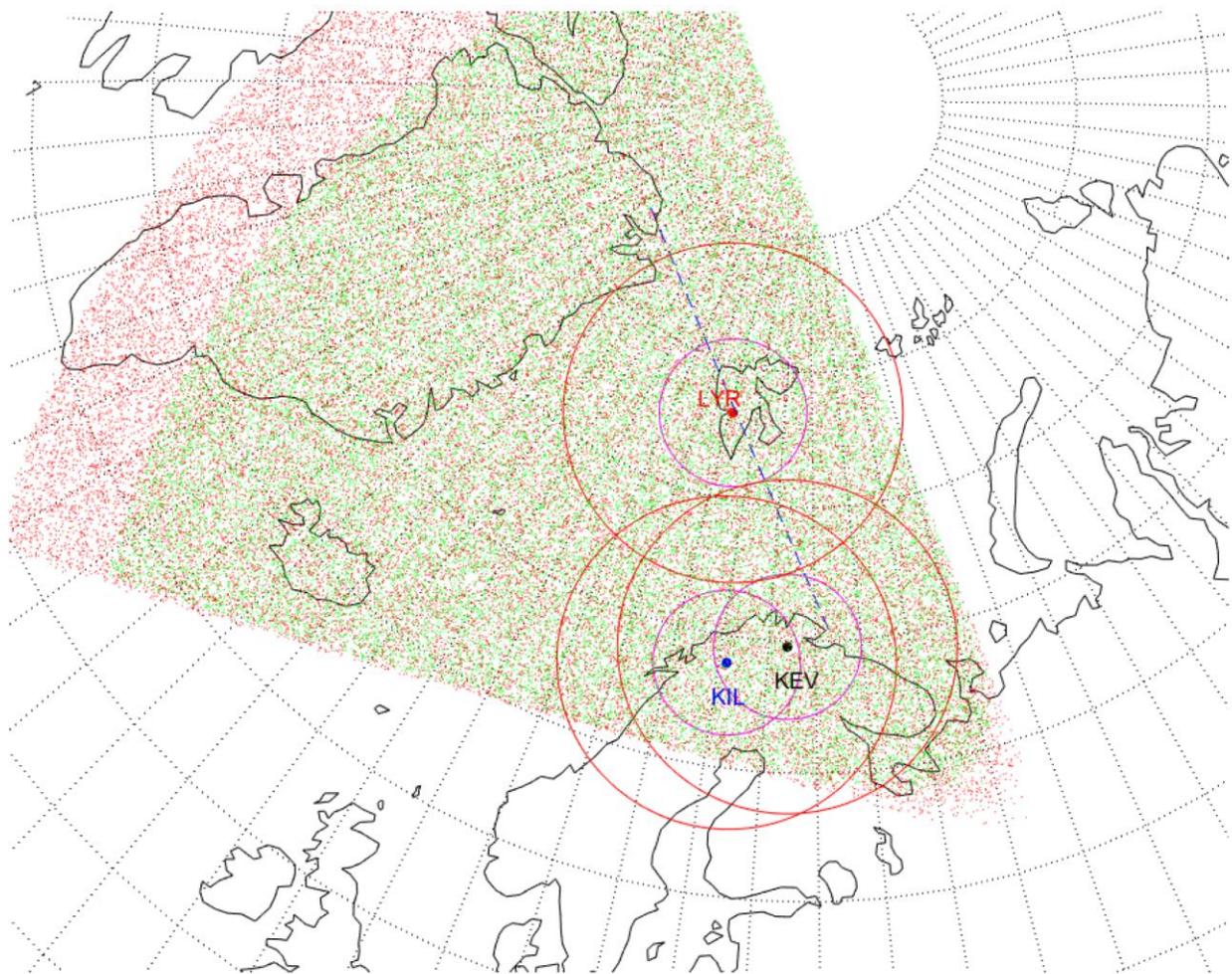

**Figure 5**. An illustration of the possible positions of auroras in the FoV of the Suomi 100 satellite's camera when the auroral image was taken. The figure shows orthographic projection viewed above Scandinavia of the simulated red and the green auroral emission positions at the altitude of 230 km and 110 km, respectively. The red, black and blue filled circles show the positions of Longyearbyen (LYR), Kevo (KEV) and Kilpisjärvi (KIL), respectively. Magenta circles around these locations display the FoVs of the all-sky cameras (ASCs). The effective FoV above the local horizon of the ASCs is assumed to be 140° (see Partamies et al., 2004, for an ASC analysis), corresponding to an area of about 600 km (magenta circles) and 1400 km (red circles) in diameter at the altitude of 110 km and 230 km, respectively. The blue dashed

line illustrates the direction of the 180°×1° FoV of Longyearbyen MSP (location adopted from Mathews et al., 2004, Fig. 1). A geographical map is shown for reference.

The modeling shown in Fig. 5 approximately estimates the maximum latitude and longitude coverage from where the auroral emission in the auroral image could have come from. However, the figure does not give an estimate for what would have been the intensity of the observed auroras in the image. This was investigated by implementing a simple virtual pin hole camera (PHC), where photon emission points from 3D space are projected on a 2D image plane and by calculating the relative intensity of the auroras.

The intensity of the auroral emission was investigated by generating red, green and blue emission source positions by using the photon emission density rate function shown in Eq. 2. Then, a perspective projection plot at the position of the Suomi 100 satellite was derived by dividing the image plane into constant size dx×dy cells, which represent virtual, relatively large, pixels on a virtual camera's image sensor. Third, the positions of the cells, [dx×dy]$_{ij}$, where the emission points are projected into the virtual image, are determined. Finally, the relative-intensity map, $I_{ij}^r$, was derived by first giving every emission point a distance-dependent intensity,

$$di_k = q \, r_k^{-2} \quad [\text{photons/s/m}^2],  \tag{5}$$

where $r_k$ is the distance between the emission point and the image plane, and $q$ [photons/s] is a normalization constant. Then the sum of all intensities within a cell (i, j)

$$di_{ij} = \sum_k^{(i,j)} di_k \quad [\text{photons/s/m}^2],  \tag{6}$$

is calculated. After that we add a front-lens model where we take into account the effective visible area of the front-lens

$$dI_{ij} = \cos \Theta_{ij} \, di_{ij} \quad [\text{photons/s/m}^2],  \tag{7}$$

where $\Theta_{ij}$ is the angle between the line-of-sight and the optical axis of the lens (see for example Gustavsson, 2000). Moreover, modeling the geometric distortions of the Suomi 100 camera was considered, but abandoned, as the distortions were too small to be detectable based on stars detected in an image of the Orion constellation and their expected image locations (c.f. Knuuttila et al., 2022). Finally, the relative intensity in the value range [0 1] is defined as

$$I_{(i,j)}^r = dI_{ij} \, / \max \, (dI_{ij}).  \tag{8}$$

Therefore, from the physical point of view, the obtained relative-intensity map simulates a situation where there is an auroral emission in 3D space in a vacuum, i.e., without the attenuation caused by the atmosphere.

The relative-intensity map is used to derive an estimation for the possible latitude and longitude positions from where the auroral emission in the auroral image could have originated. This was accomplished by pointing the virtual camera in the direction where the aurora was seen in the auroral image and using $50×10^6$ red, green, and blue emission source positions. The emission positions were collected within the simulated pixels of the virtual image approximately in the same place where the real aurora exists in the auroral image. In Fig. 6, the region of the aurora is marked in the simulated auroral emission image as a blue square. As for a quantitative estimate of the collected auroral emission, the red, green and blue relative-intensity maps contain 871 527, 644 249 and 651 188 auroral emission points, respectively, i.e., the ratio of the total number of the collected emission positions of the red/green/blue aurora was 1/0.74/0.75. In Fig. 6 all simulated emissions are normalized by their maximum intensity. The ratios of the maximum intensity of the simulated red($\max(I_{630.0\ nm})$), green($\max(I_{557.1})$) and the blue($\max(I_{427.8\ nm})$) emission were 1/1.53/0.29. This implies that the highest maximum intensity was in the green channel and the smallest maximum intensity was in the blue channel.

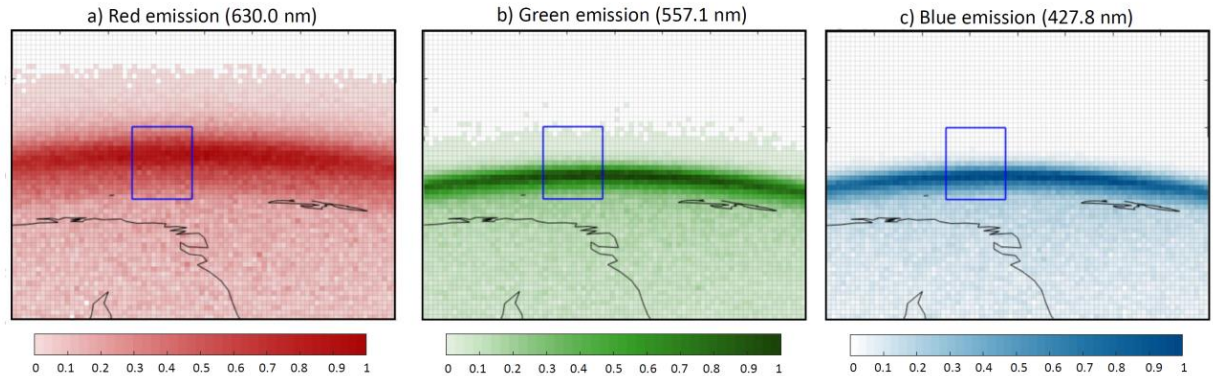

**Figure 6**. Simulated relative intensities $I^r_{630.0\ nm}$, $I^r_{557.1\ nm}$, and $I^r_{427.8\ nm}$ from the red (a), green (b) and blue (c) auroral emission, respectively. Each intensity is normalized to its maximum value. The blue squares show the approximate center position of the auroras in the auroral image. The maps show the perspective projection plots. The maximum altitude of the red, green, and blue auroras was assumed to be 250 km, 110 km, and 110 km, respectively. The coastal lines are shown for reference.

Fig. 6 shows how the position of the highest intensity of the red auroras in the simulated image is higher from the limb of the Earth than that of the green and blue auroras. As already noted, some inaccuracy to the simulated values occurs because the model does not take into account atmospheric attenuation, which is strongest near the local horizon where the atmospheric densities are the highest (see e.g., Scheidegger et al., 2008). The essential information obtained from the simulated relative-intensity maps shown in Fig. 6, however, is that the clearest auroras can be detected near the limb of the Earth.

Although Fig. 6 shows the intensities in three emission wavelengths, $I_{630.0\ nm}$, $I_{557.1\ nm}$, and $I_{427.8\ nm}$, they cannot be directly compared with the observed counts in the red($R_{obs}$), green($G_{obs}$) and blue($B_{obs}$) channels in the actual auroral image because of several camera effects. First, if the lens transmittance varies noticeable in the analyzed wavelength range, the simulated intensities should be calibrated by using the transmittance. The analyzed case variation of the transmittance between ~400 nm – 800 nm is, however, so small that it can be approximated to be a constant (see Knuuttila et al., 2022, Fig. 3a). Second, the camera's color channels collect light from a wide range of wavelengths. For example, the camera's R-channel collects photons from the red (630 nm), green (557.1 nm) and blue (427.8) wavelengths, although the red channel is most sensitive in these wavelengths to 630 nm. Therefore, counts in the simulated R($R_{sim}$), G($G_{sim}$) and B($B_{sim}$) channels are derived by using the camera's quantum efficiencies, $c^R_l$, $c^G_l$, and $c^B_l$, where $\lambda$ is the simulated wavelength:

$$R_{sim} = c^R_{630.0\ nm}\, I_{630.0\ nm} + c^R_{557.1\ nm}\, I_{557.1\ nm} + c^R_{427.8\ nm}\, I_{427.8\ nm}, \qquad (9a)$$

$$G_{sim} = c^{G}_{630.0\,nm}\, I_{630.0\,nm} + c^{G}_{557.1\,nm}\, I_{557.1\,nm} + c^{G}_{427.8\,nm}\, I_{427.8\,nm}, \qquad (9b)$$

$$B_{sim} = c^{B}_{630.0\,nm}\, I_{630.0\,nm} + c^{B}_{557.1\,nm}\, I_{557.1\,nm} + c^{B}_{427.8\,nm}\, I_{427.8\,nm}. \qquad (9c)$$

The values of the quantum efficiencies of the different channels at the wavelengths of interest are estimated from the camera's datasheet (see Knuuttila et al., 2022, Fig. 3b). The adopted values are $[c^{R}_{630.0\,nm}, c^{R}_{557.1\,nm}, c^{R}_{427.8\,nm}] = [34, 7, 2]$ %,  $[c^{G}_{630.0\,nm}, c^{G}_{557.1\,nm}, c^{G}_{427.8\,nm}] = [8, 32, 3]$ %, and $[c^{B}_{630.0\,nm}, c^{B}_{557.1\,nm}, c^{B}_{427.8\,nm}] = [4, 7, 26]$ %.

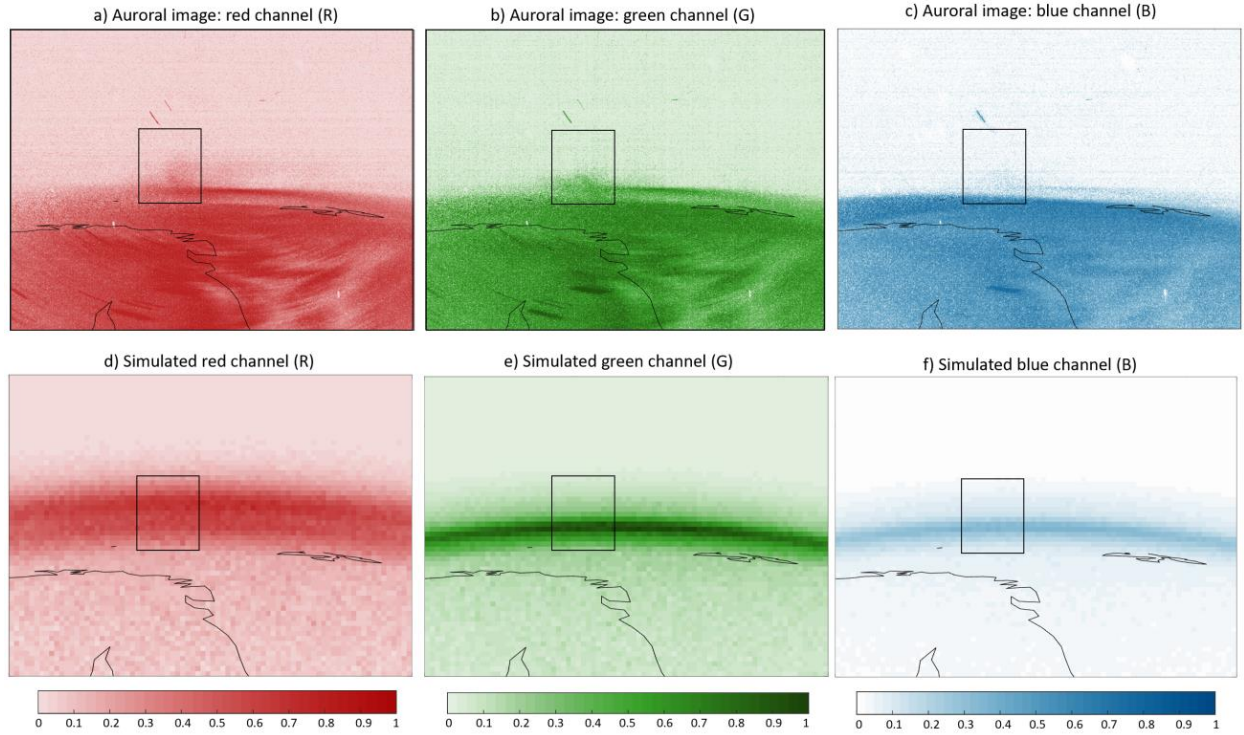

**Figure 7**. Comparison of the auroral image (a-c) taken on January 22$^{nd}$, 2019, with the simulated image (d-e). Panels a), b) and c) show the normalized values in the observed red($R_{obs}$), green($G_{obs}$) and blue($B_{obs}$) color channels of the auroral image. The maximum count in the red, green, and blue channels were 253, 217 and 255, respectively. Panels d), e) and f) show the simulated normalized red($R_{sim}$), green($G_{sim}$) and blue($B_{sim}$) color channels, respectively. The black squares show the approximate center position of the observed auroras. The maps show the perspective projection plots. The maximum altitude of the red, green, and blue auroras was assumed to be 230 km, 110 km, and 110 km, respectively. The coastal lines are shown for reference.

Comparison between the observed and simulated counts in the red, green, and blue channels are shown in Fig. 7. One can make the following observations. First, the auroral image's blue channel (Fig. 7c) does not show as strong an auroral emission as the red and green channels. The same is true also for the simulated color channels because the ratio of the maximum simulated red, green, and blue channels, $\max(R_{sim})/\max(G_{sim})/\max(B_{sim})$, is $1/1.51/0.55$ that is, the maximum simulated green channel counts are the largest. However, note that the $\max(G_{sim})/\max(B_{sim})$ count ratio is only 2.7, that is, it is smaller than the emission ratio of the green and blue auroras (= 5) in the model, because the G channel is sensitive also to red and green light (Eq 9c) which has high intensities. Second, both the observed (Fig. 7a) and the simulated (Fig. 7d) red channel gives far more vertically spread auroral emission what is seen in the green and blue channels. The increased "diffusion" of the aurora in the simulated red channel compared with the green and blue channels is caused by the larger scale height of the red auroral emission profile (50 km) compared with the scale height of the green and blue emission profiles (25 km). The aurora in the simulated red channel is also higher up from the horizon than auroras seen in the green and blue channels because the peak emission altitude of the simulated red aurora (230 km) was at a higher attitude than the peak emission altitude of the simulated green and blue auroras (110 km).

The positions of the red and green emission points within the black squares in Fig. 7 are then collected and shown in Fig. 8. Note that the emission positions of the blue aurora are not shown in Fig. 8 because it has the same spatial distribution as the blue aurora since the shape of their Chapman functions are identical. This figure shows that those auroral emissions could have originated from the Norwegian Sea and above Greenland. However, as seen earlier in Fig. 5, the green auroras must have been somewhere close to a center of a triangle with corners at Longyearbyen, the northern part of Scandinavia and at Iceland because: (i) magnetometers in Greenland did not show magnetic activity, and (ii) it was not seen in the ASCs at Longyearbyen, Kevo and Kilpisjärvi. Therefore, the auroras were also outside of the FoV of the Longyearbyen MSP, during the time of the auroral image.

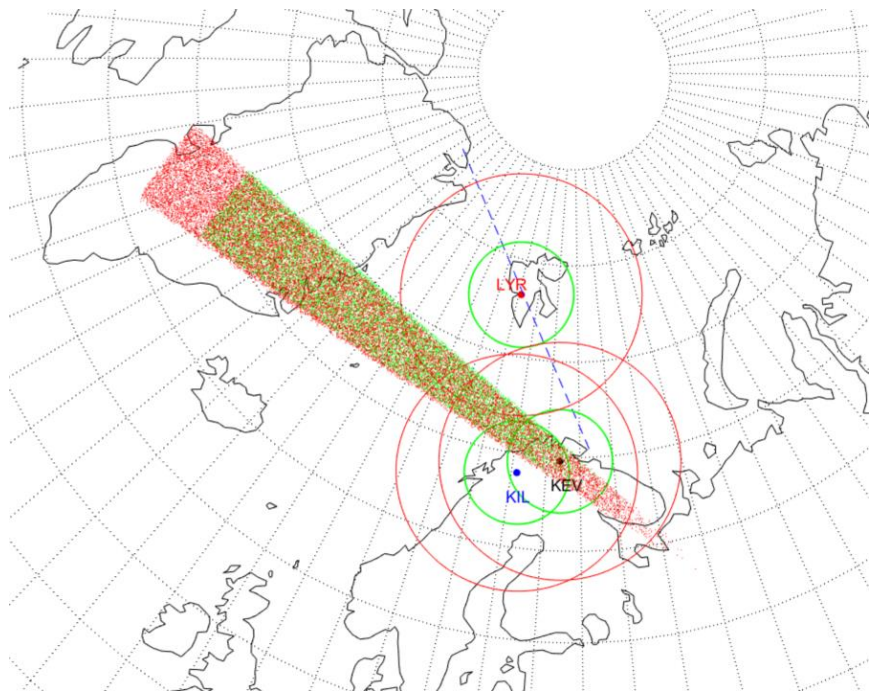

**Figure 8**. Possible positions of auroral emission seen in the auroral image. The green and the red points show the position of auroral emission within the black squares shown in Fig. 7. The green and red circles show the FoV where ASCs can see green and red auroras at altitudes 110 km and 230 km, respectively, similar as in Fig. 5. The blue dashed line represents the direction of the FoV of Longyearbyen MSP similar to as in Fig. 5. The red, black, and blue filled circles show the positions of Longyearbyen (LYR), Kevo (KEV) and Kilpisjärvi (KIL), respectively. The position of auroras is shown in an orthographic projection.

One can argue that the most realistic model for a discrete aurora might be if the auroral emission comes near some constant geomagnetic latitude. That situation was investigated by adding emission positions around a constant geomagnetic latitude (Fig. 9). In the model, the geomagnetic latitude was taken to be 71.58°N, that is, close the geomagnetic latitude of the Bear Island station. The emission points were located smoothly around the central latitude using a normal distribution with a standard deviation of 0.25°. The longitudes of the emission points were sampled uniformly from the range [110° 115°], which was chosen so that the emission location in the simulated image matches the aurora location in the auroral image.

As a result, as can be seen in Fig. 9a, the discrete aurora was in this model near the edge of the FoVs of the ASCs where they cannot detect any green auroral emissions and where they can only detect the red auroral emission with difficulty. Note also that the vertical distance of the peaks of the red and green emissions are different than in Fig. 7. The difference is due to the larger distance between the auroral emission and the virtual camera.

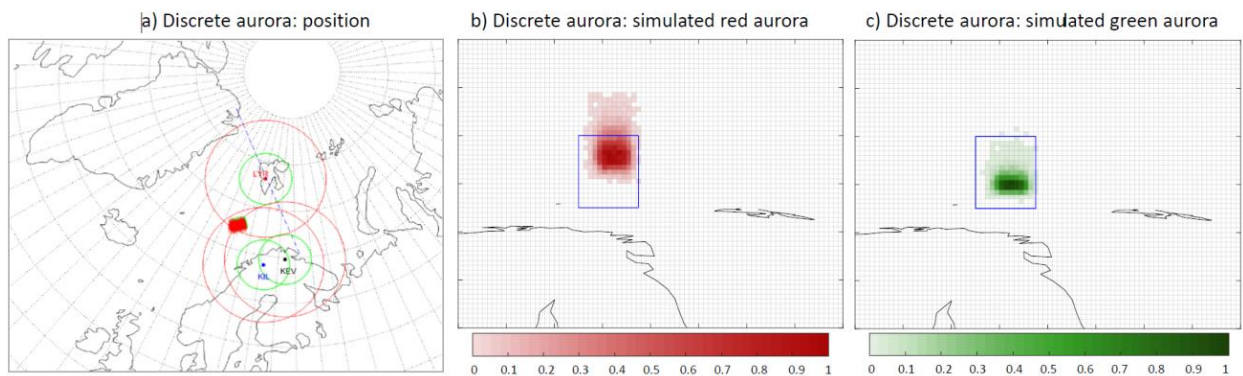

**Figure 9**. The discrete aurora model where the emission was assumed to be near a constant geomagnetic latitude (71.58° N) in a narrow longitude range ([110° 115°]). In panel a) the red and green dots show the auroral emission positions. Panels b) and c) show the simulated normalized emission intensities, similar as show in Fig. 6.

## 5 The limb geometry effect

The analysis of the auroral image raises several important questions about the potential of a satellite on an SSO with a camera instrument to investigate auroras. Especially, one can ask how a wide-angle ASC on a satellite would have seen the auroras in the auroral image case? And especially, how is the imaged auroral emission affected by the limb geometry?

Fig. 6 suggested that a favorable direction to observe intense auroral emission would be the limb of the Earth. However, Fig. 6 is affected by the different solid angles subtended by the PHC pixels. Particularly, in Fig. 6 the relative intensity depends on the direction of the camera plane because constant size cartesian pixels subtend a different area on the unit sphere when using the perspective projection. Therefore, for instance, the pixels near the center of the image receive more light than the pixels further away from the center. This affects the interpretation of the image as the emission would appear brighter at the center than at the edges.

This camera direction effect can be removed, when the intensity at every virtual pixel (i, j), is divided by its solid angle, $d\Omega_{ij}$, i.e.

$$di_{ij,\Omega} \equiv di_{ij}/d\Omega_{ij} \qquad [\text{photons/s/m}^2/\text{str}], \qquad (10)$$

where $di_{ij}$ is the sum of all intensities within a cell (i, j) (c.f. Eq. 6). The $d\Omega_{ij}$ at every pixel on the image plane is calculated from the positions of the four corners of the pixel (see e.g., Mazonka, 2012). Therefore, $di_{ij,\Omega}$ includes the geometric attenuation of the light ($\sim 1/r^2$) but not the front-lens effect ($\sim \cos\Theta_{ij}$). The resulting directional differential flux ($di_{ij,\Omega}$), is then normalized by dividing its value by the maximum value on the image plane giving the normalized directional differential flux, $d\tilde{\imath}_{ij,\Omega}$.

An auroral emission model was made which included $50\times10^6$ green and red emission sources. A 150° FoV PHC at an altitude of 600 km looked to the nadir. The model showed that in this situation the camera would have seen $\sim 5.5\times10^6$ green and red emission sources, that is, $\sim$11% of the emission points. Taking into account the geometric attenuation of the light ($\sim 1/r^2$) and the fact that every red emission point emits about the same number of photons than every green emission point ($w_r/w_g = 1.0373$), the ratio of the collected total number of simulated red photons and the total number of simulated green photons is $\sim$1, that is, the total amount of red and green light is quite similar. Moreover, the ratio of the collected total number of simulated blue photons and the total number of simulated green photons is 0.2 because the spatial distribution of their emission locations are identical, while every blue emission point emits only 20% of photons that every green aurora source point emits ($w_b/w_g = 0.2$).

Fig. 10 shows the resulting normalized directional differential flux, $d\tilde{\imath}_{ij,\Omega}$, for the red and green auroras. The blue aurora image is not shown because it would look similar to the green emission aurora, the only difference being that the emission would be 20% of the green auroral emission. Note that the limb of the Earth spans around 130°, i.e., the whole surface of the Earth seen at that altitude could have already been imaged with a 130° FoV camera. Moreover, imaging the peak of the modeled red auroral emission at an altitude of 230 km would be possible with a $\sim$140° FoV camera, which would capture the Earth and a 250 km thick spherical shell around it.

There are several notable features in Fig. 10. First, due to perspective projection, the emission points close to the camera form a thick emission layer above the local horizon. This is especially true for the red aurora because its emission layer altitude is closer to the camera than the green aurora layer.

The most important information obtained from Fig. 10 is that the highest relative intensities can be seen near the local horizon, i.e., above the limb of the Earth. The brightness in this direction occurs because the look direction is along the side of the emission layer where the emissions points are "packed" close to each other when viewed from the satellite. Therefore, it is not unexpected that Suomi 100's camera observed auroras in the auroral image while looking toward the limb of the Earth, because it is the direction which is the most favorable imaging direction to detect high auroral emission intensities.

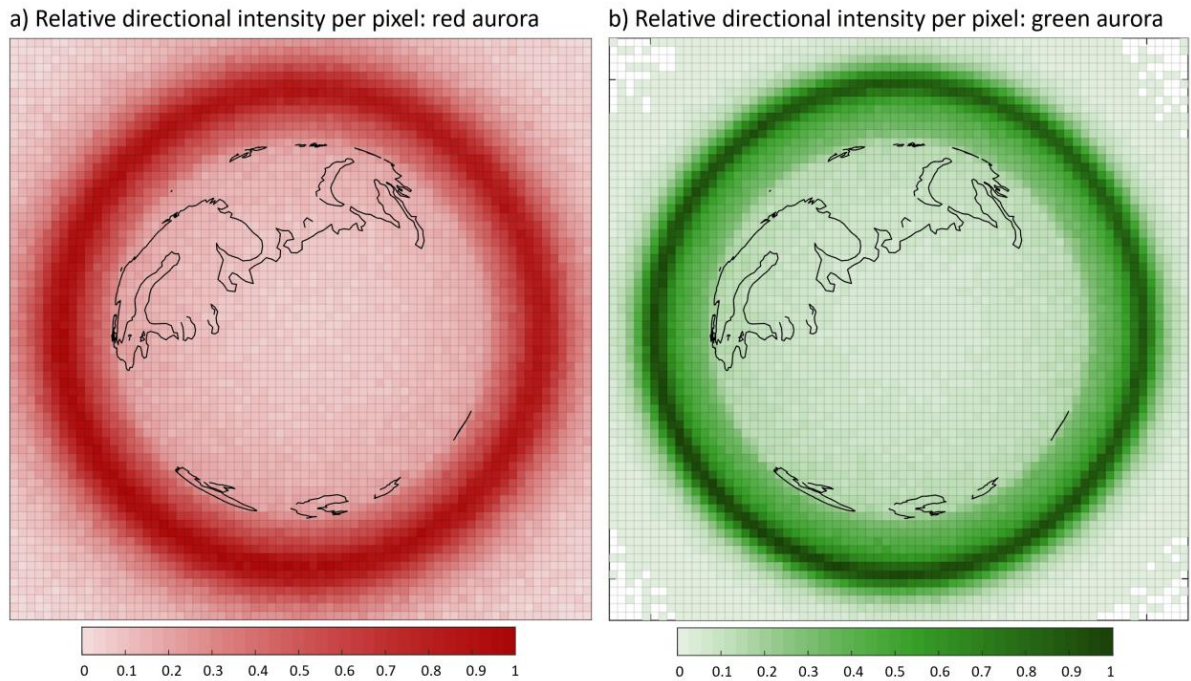

**Figure 10**. A simulated normalized directional differential flux ($d\tilde{\imath}_{ij,\Omega}$) maps of the modeled (a) red and (b) green auroras seen by a virtual 150° FoV camera on a satellite at the position where the auroral image was taken. The color shows the relative intensity. The altitude of the green and red auroras was modeled to be 110 km and 230 km, respectively. Both green and red intensity maps are normalized individually, the maximum green emission being 1.36 times larger than the maximum red emission. A geographical map is shown for reference.

## 6 Discussion and concluding remarks

This paper presents a multi-instrument case study where optical observations from a nanosatellite on January 22$^{nd}$, 2019, are analyzed alongside a ground-based auroral camera and magnetometer observations. The auroral imaging took place on the Suomi 100 satellite mission's camera operation phase in December 2018 – February 2019, which provided good dark auroral imaging conditions over Northern Europe. The analyzed image is, to the authors' best knowledge, the first auroral image ever taken by a cubesat. After the 1$^{st}$ science period, the satellite started its 2$^{nd}$ science mission, which focused on performing measurements with its radio spectrometer instrument (Kallio et al., 2022).

The estimation of the probable origin of the observed auroral emission in the analyzed auroral image showed that the aurora could have been located over the North Sea or above Greenland. The fact that ASCs in Svalbard, Kevo and Kilpisjärvi did not, however, observe auroras imply that the emission was located outside the FoV of these cameras. If that is the case, this is an example of a situation where space-based imaging was successful while ground-based imaging was not. The model of the relative intensities showed that the pointing direction of the Suomi 100's camera during the exposure of the auroral image, which was toward the limb of the Earth, was highly favorable to observe faint auroras. Our analysis demonstrates that the capabilities of spacecraft auroral imaging are very different to ASCs and, consequently, that images from even a nanosatellite can fill the gaps existing in data obtained from ground-based ASCs alone.

The position of the aurora was investigated by using 3D auroral position and virtual camera models. The approach used here to generate a simulated auroral image has similarities with the modeling of energetic neutral atom (ENA) observations in space: simulated particles, in this case photons, were generated randomly and the properties of the particles collected by a virtual instrument, which depends on the relative position of the source region and a virtual camera, provides physical, line-of-sight integrated, information about the investigated region (see Kallio et al., 2006). The emission altitude profiles were assumed to be simple Chapman production functions for simplicity. However, in the future, more detailed emission profiles could also be used such as provided by the ionospheric codes GLOW (Solomon, 2017) and Transsolo (Lilensten et al., 2015). For example, in this work the emission profile shapes of the green and blue lines were assumed to be exactly identical for simplicity (see Whiter et al., 2023, for a statistical analysis of the altitude of green OI 557.7 nm and blue $N_2^+$ 427.8 nm aurora).

Synthetic images based on the intensity model showed that the location where the auroras were observed were ideal for auroral photography from space, because the integrated intensity of the emission from the aurora is highest near the local horizon of the spacecraft. This kind of limb-pointing measurement is ideal also because the background of the aurora on the nightside is black sky. The auroral emission from the nadir could, instead, be easily mixed with the artificial light from the ground and the scattering of light from the clouds, which was also the case in part of the analyzed auroral image. It is interesting to note that, although it would be challenging to include and operate a 180° FoV camera in a satellite, the FoV of even a typical spacecraft camera can be increased if the satellite rotates slowly. Then several photographs can be combined into a panoramic image of the auroral region similarly to the daylight panoramic image (Fig. 2). However, if the satellite is large, then also dedicated camera solutions are possible. For example, the wide-field auroral imager (WAI) on the Fengyun-3D satellite has two cameras which provide in total 130° × 10° FoV (Zhang et al., 2019).

It should be noted that although the developed relatively simple 3D auroral model gave insight for the position of the auroral emission and the relative intensity obtained from different directions, the final step in the auroral reconstruction model is to derive the differential directional energy flux, or particle flux, in 3D space from auroral images. However, making a reconstruction model which provides estimation of the electron precipitation flux is a highly challenging inversion task (see for example, Solomon et al., 1984; Solomon, 1987; Janhunen, 2001; Simon-Wedlund et al., 2013, and discussion therein). Another modeling approach may also be to use a model that describes the emission from the precipitation of a given particle species, for example, associated with the precipitating protons (see e.g., Kallio & Janhunen, 2001). Although this kind of challenging reconstruction approach is beyond the scope of the present study, one can anticipate that photographs taken by satellites would provide help for such detailed inversion tasks because satellite images give different line-of-sights than images obtained from ground-based ASCs.

Although the presented study does not provide a 3D tomographic construction of the observed aurora, the study suggests that differences between the positions of the maximum green and red auroral emission in the RGB image may be used to estimate the distance of a discrete aurora from the camera (c.f. Fig. 9). This question was further investigated by the generated aurora RGB images where the discrete emission was modeled to be within a certain distance range from the camera, dr = 250 km, and the center of the aurora being 2000 km, 3000 km, and 3600 km from the camera. The analysis showed that the greater the distance of the aurora from the camera, the larger the vertical distance of the maximum red and green auroral emission positions was in the figure. Therefore, even a single image can provide useful information about the position of a discrete aurora. However, in the analyzed auroral image the red, green, and blue emissions were so faint that this kind of distance analysis could not be done accurately. Moreover, the analyzed auroral image taken at 18:08:13 UT is the only full-size image of the investigated aurora, and all other images of the event by the Suomi 100 satellite are thumbnails, which do not give a possibility to compare images of the aurora from different directions. It is also worth noting that even a more detailed analysis of the position of a discrete aurora could be obtained if a long-living discrete aurora could be imaged at two or more positions along the orbit of the spacecraft (c.f. Fig. 3).

One can also anticipate that a constellation of cubesats with auroral cameras would increase possibilities for a more accurate auroral position analysis because auroras could be imaged at the same time from different directions. Moreover, instead of a typical RGB camera, an RGBW image sensor with additional higher sensitivity panchromatic (W) pixels, as used in the AMICal Sat cubesat's camera (Barthelemy et al., 2022), together with a pass-band optical filter, could increase the possibility of capturing faint auroral emissions and help resolve the emission intensities at different wavelengths. However, the image sensor used for the AMICal camera mainly consists of panchromatic pixels with only one colored pixel in each 4×4 pixel group and, at the time of writing, it seems that there is only a very limited selection of RGBW sensors in the market.

It should also be noted that the downlink data rate can be a limiting factor in scientific investigations using auroral imaging by a cubesat and higher rates would provide more possibilities. For example, the Suomi 100 satellite uses low data rate VHF frequency (437.775 MHz) and, therefore, most of the downlinked images so far have been thumbnails. Together with better image compression using e.g. JPEG2000, one can envision that modern pattern recognition and machine learning algorithms provide useful possibilities for such a low downlink data rate space mission. In the Suomi 100 satellite case, the choice of which full-sized images are downlinked has so far been based on manual inspection of the thumbnails. However, , software was developed before the launch to analyze the images already on board the satellite. This analysis software included three different algorithms which were designed to evaluate the likelihood of having auroras in an image based on the image HSV histogram (Hue, Saturation and Value), calculated from its RGB values. The developed low-computational cost algorithms were tested using ground-based ASC images of auroras. However, together with the required short exposure time – partly due to the spinning of the satellite – and the relatively low geomagnetic activity, imaging has so far resulted in auroral intensities that are too weak for the algorithms to identify unambiguously. Therefore, the selection of which full-size images to downlink has so far been performed manually.

It is noteworthty that the analyzed auroral image was taken during the northern hemispheric winter when the ambient light conditions were ideal for auroral photography. However, the first two-and-a-half years of the Suomi 100 satellite's mission from approximately December 2018 until July 2022, i.e., about the first 20,000 orbits, have not so far been ideal for auroral observations because of the Sun's low activity. The solar activity reached a minimum between the end of Solar Cycle 24 and the start of Solar Cycle 25 in December 2019. Fortunately, the satellite is estimated to remain in orbit for over ten years during which time the solar activity, and thus auroral activity, will increase on average.

The presented analysis showed basic pros and cons of using a small satellite for auroral imaging. Small-satellite observations benefit significantly from the combination with ground-based observations. In such multi-instrument studies, the satellite observations can either provide new or complementary observations. Ground-based optical data is not available for the case demonstrated here, but we foresee the combination of nanosatellite and ground-based cameras as a useful step forward in the tomographical imaging of auroras. The challenge of a small satellite is that the simplicity of its instruments and the difficulties obtaining a high pointing accuracy of the camera complicates the data analysis. On the other hand, the attitude of a small satellite and the operation of the instruments can typically be controlled more freely than in the case of a large, expensive multi-instrument satellite. This case study is the first time a simple camera instrument on a small satellite is used in auroral research in combination with ground-based observations and modeling.

## Acknowledgments

The authors would like to thank the Finnish Prime Minister's Office, The Magnus Ehrnrooth Foundation and the Academy of Finland (Decision No. 348784) for the financial support of the Suomi 100 satellite project. Tuija Pulkkinen is acknowledged for her support of the Suomi 100 satellite project. EK thanks Jari Mäkinen for the Suomi 100 satellite outreach activities and Valtteri Harmainen for the design and implementing of the data-sharing cloud service. The panoramic image in Figure 2 is made with the Hugin open-source panoramic photo-stitching and merging program (http://hugin.sourceforge.net/). The authors acknowledge Dr. Max van de Kamp for the preparation of Fig. 4c and the Finnish Meteorological Institute for the used 2D Equivalent Currents web service (https://space.fmi.fi/MIRACLE/iono_2D.php).

## Open Research

Auroral electrojet activity can be obtained from https://space.fmi.fi/MIRACLE/iono_2D.php, ASC keogram from Sodankylä from https://space.fmi.fi/MIRACLE/ASC/?page=keograms, Sodankylä Geomagnetic Observatory's riometer observations from https://www.sgo.fi/Data/Riometer/rioData.php, and the IMAGE magnetometer data from https://space.fmi.fi/image/www/. The original auroral image (img001229.jpg) and the polished auroral image (img001229_processed.jpg) that has been used in this study are available at Kallio, 2023.

## References

Amm, O., and A. Viljanen, A. (1999). Ionospheric disturbance magnetic field continuation from the ground to the ionosphere using spherical elementary current systems, *Earth Planets Space*, 51(6), 431–440. https://doi.org/10.1186/BF03352247

Barthelemy, M., Kalegaev, V., Vialatte, A., Le Coarer, E., Kerstel, E., Basaev, A., et al. (2018). AMICal Sat and ATISE: two space missions for auroral monitoring, *J. Space Weather Space Clim*., 8, A44. https://doi.org/10.1051/swsc/2018035

Barthelemy, M., Robert, E., Kalegaev, V., Grennerat, V., Sequies, T., Bourdarot, G., et al. (2022). AMICal Sat: A sparse RGB imager on board a 2U cubesat to study the aurora, IEEE Journal on Miniaturization for Air and Space Systems, Vol. 3, No. 2.

Brändström, U. (2003). The Auroral Large Imaging System - Design, Operation and Scientific Results, IRF Scientific Report 279, ISBN 91-7305-405-4, Kiruna, Sweden.

Dreyer, J., Partamies, N., Whiter, D., Ellingsen, P. G., Baddeley, L., and Buchert, S. C. (2021). Characteristics of fragmented aurora-like emissions (FAEs) observed on Svalbard, *Ann. Geophys*., 39, 277–288. https://doi.org/10.5194/angeo-39-277-2021

Ellingsen, P. G., Lorentzen, D., Kenward, D., Hecht, J. H., Evans, J. S., Sigernes, F., and Lessard, M. (2021). Observations of sunlit N2+ aurora at high altitudes during the RENU2 flight, *Ann. Geophys*., 39, 849–859. https://doi.org/10.5194/angeo-39-849-2021

Enell, C.-F. , B. Gustavsson, B. U. E. Brändström, T. I. Sergienko, P. T. Verronen, P. Rydesäter, and I. Sandahl (2012). Tomography-like retrieval of auroral volume emission ratios for the 31 January 2008 Hotel Payload 2 event, *Geosci. Instrum. Method. Data Syst*., 2, 1–21. https://doi.org/10.5194/gid-2-1-2012

Gustavsson, B. (2000). Three dimensional imaging of aurora and airglow, PhD thesis, Swedish Institute of Space Physics, Kiruna, Sweden.

Jackel, B. J., Creutzberg, F., Donovan, E. F., and Cogger L. L. (2003), Triangulation of Auroral Red-Line Emission Heights, *Sodankylä Geophysical Observatory Publications*, 92:97–100. https://aurora.phys.ucalgary.ca/donovan/pdfs/jackel_triangulation_oulu_2003.pdf

Janhunen, P. (2001). Reconstruction of electron precipitation characteristics from a set of multiwavelength digital all-sky auroral images. *Journal of Geophysical Research: Space Physics*, Vol. 106, IA9, 18505-18516. https://doi.org/10.1029/2000JA000263

Kallio, E. (2023). Suomi 100 satellite's images: The auroral and the star images [Dataset]. Zenodo. https://doi.org/10.5281/zenodo.7645289

Kallio, E., Kero, A., Harri, A.-M., Kestilä, A., Aikio, A., Fontell M., et al. (2022), Radar – CubeSat Transionospheric HF Propagation Observations: Suomi 100 Satellite and EISCAT HF Facility, *Radio Science*. Vol 57, No 10. https://doi.org/10.1029/2022RS007516

Kallio, E. et al. (2006). Energetic Neutral Atoms (ENA) at Mars: Properties of the hydrogen atoms produced upstream of the martian bow shock and implications for ENA sounding technique around non-magnetized planets, *Icarus*, Vol. 182, Issue 2, 448-463. https://doi.org/10.1016/j.icarus.2005.12.019

Kallio, E. and Janhunen, P. (2001). Atmospheric effects of precipitating energetic hydrogen atoms on the Martian atmosphere, *Journal of Geophysical Research: Space Physics*,Vol 106, A1, 165-177. https://doi.org/10.1029/2000JA002003

Knuuttila, O., Kallio, E. Partamies, N., Syrjäsuo, M., Kauristie, K., Sofieva, V., et al. (2022). In-space Calibration of Nanosatellite Camera, *The Journal of Small Satellites. Vol. 11, No3*.

Lilensten, J., Bommier, V., Barthélemy, M., Lamy, H., Bernard, D., Moen, J. I., Johnsen, M. G., Løvhaug U. P., and Frédéric Pitout, P. (2015). The auroral red line polarisation: modelling and measurements, J. Space Weather Space Clim., 5, A26. https://doi.org/10.1051/swsc/2015027

Mathews, J. T., Mann, I. R., Rae, I. J., and J. Moen (2004). Multi-instrument observations of ULF wave-driven discrete auroral arcs propagating sunward and equatorward from the poleward boundary of the duskside auroral oval, *Physics of Plasmas*, Vol. 11, No 4. https://aip.scitation.org/doi/10.1063/1.1647137

Mende, S. B., Harris, S. E., Frey, H.U., Angelopoulos, V., Russel, C.T., Donovan, E., et al. (2009). The THEMIS Array of Ground-based Observatories for the Study of Auroral Substorms. In: Burch, J.L., Angelopoulos, V. (eds) *The THEMIS Mission*, Springer, New York, NY. https://doi.org/10.1007/978-0-387-89820-9_16

Moore C. S., Caspi, A., Woods, T. N., Chamberlin, P. C., Dennis, B. R., Jones, A. R., et al. (2018). The Instruments and Capabilities of the Miniature X-Ray Solar Spectrometer (MinXSS) CubeSats, *Solar Phys*, 293:21. https://doi.org/10.1007/s11207-018-1243-3

Mazonka, O. (2012). Solid Angle of Conical Surfaces, Polyhedral Cones, and Intersecting Spherical Caps, *arXiv*. https://doi.org/10.48550/arXiv.1205.1396

Ogawa, Y., Tanaka, Y., Kadokura, A., Hosokawa, K., Ebihara, Y., Motoba, T., et al. (2020). Development of low-cost multi-wavelength imager system for studies of aurora and airglow, *Polar Science*, 23, 100501. https://doi.org/10.1016/j.polar.2019.100501

Oikarinen, L. (2001). Polarization of light in UV-visible limb radiance measurements, *Journal of Geophysical Research*, 106, 1533–1544. https://doi.org/10.1029/2000JD900442

Partamies N., Janhunen, P., Kauristie, K., Mäkinen, S. and Sergienko, T. (2004). Testing an inversion method for estimating electron energy fluxes from all-sky camera images, *Annales Geophysicae*, Vol. 22, 1961–1971. Ann. Geophys., 22, 1961–1971. https://doi.org/10.5194/angeo-22-1961-2004

Partamies, N., Syrjäsuo, M., & Donovan, E. (2007). Using colour in auroral imaging. Canadian         journal         of         physics,         85(2),         101-109. https://doi.org/10.1139/p06-090

Partamies, N., Whiter, D., Syrjäsuo, M., and Kauristie, K. (2014). Solar cycle and diurnal dependence of auroral structures, *Journal of Geophysical Research: Space Physics*, Vol. 119. https://doi.org/10.1002/2013JA019631

Qiu Qi, Yang Hui-Gen, Lu Quan-Ming, and Hu Ze-Jun (2017). Correlation between emission intensities in dayside auroral arcs and precipitating electron spectra, *Chinese Journal of Geophysics*, 60(1): 1-11. http://html.rhhz.net/Geophy_en/html/20170101.htm

Rossi, S., Ivanov, A., Richards, M., and Gass, V. (2013). The SwissCube's technologies results after four years of Flight, Conference: *International Astronautical Conference*, Paper number: IAC-13, B4,6B,5,x16621, Bejing.

Sangalli, L., Partamies, N., Syrjäsuo, M., Enell, C.-F., Kauristie, K., and Mäkinen S. (2011). Performance study of the new EMCCD-based all-sky cameras for auroral imaging, International *Journal of Remote Sensing*, Vol. 32, 2987-3003. https://doi.org/10.1080/01431161.2010.541505

Scheidegger, N., Shea, H., Charbon, E., and Rugi-Grond, E. (2008). Low Cost Earth Sensor Based on Oxygen Airglow (AIRES), D8 – Final Report, 2008 EPFL & Oerlikon Space, https://infoscience.epfl.ch/record/126161#record-files-collapse-header

Shiokawa, K., Katoh, Y., Hamaguchi, Y., Yamamoto, Y., Adachi, T., Ozaki, M., et al. (2017). Ground-based instruments of the PWING project to investigate dynamics of the inner magnetosphere at subauroral latitudes as a part of the ERG-ground coordinated observation network, *Earth, Planets and Space*, 69, 160. https://doi.org/10.1186/s40623-017-0745-9

Shiokawa, K., Otsuka, Y., and Connors, M. (2019). Statistical study of auroral/resonant-scattering 427.8-nm emission observed at subauroral latitudes over 14 years, *Journal of Geophysical Research: Space Physics*, Vol. 124, 9293–9301. https://doi.org/10.1029/2019JA026704

Simon Wedlund, C., H. Lamy, B. Gustavsson, T. Sergienko, and U. Brändström (2013), Estimating energy spectra of electron precipitation above auroral arcs from ground-based observations with radar and optics, J. Geophys. Res. Space Physics, 118, 3672–3691. https://doi.org/10.1002/jgra.50347

Solomon, S.C., Hays, P. B., and Abreu, V. J. (1984). Tomographic inversion of satellite photometry, *Applied Optics*, 23, Issue 19, pp. 3409-3414. https://doi.org/10.1364/AO.23.003409

Solomon, S. C. (2017). Global modeling of thermospheric airglow in the far ultraviolet, J. Geophys. Res. Space Physics, 122, 7834–7848. https://doi.org/10.1002/2017JA024314

Solomon, S. C. (1987). Tomographic inversion of auroral emissions, *University of Michigan ProQuest Dissertations Publishing, ProQuest Dissertations Publishing*, 1987. 8712213, ISBN979-8-206-00988-0. https://www.proquest.com/dissertations-theses/tomographic-inversion-auroral-emissions/docview/303570931/se-2

Syrjäsuo, M., (2001). FMI All-Sky Camera Network, *Geophysical Publications* Nr. 52, ISBN 951-697-543-7.

Tanskanen, E. I. (2009). A comprehensive high-throughput analysis of substorms observed by IMAGE magnetometer network: Years 1993–2003 examined, *Journal of Geophysical Research*, Vol. 114, A05204. https://doi.org/10.1029/2008JA013682

Whiter, D. K., Partamies, N., Gustavsson, B., and Kauristie, K. (2023). The altitude of green OI 557.7 nm and blue $N_2^+$ 427.8 nm aurora, Ann. Geophys., 41, 1–12. https://doi.org/10.5194/angeo-41-1-2023

Whiter D. K., Gustavsson, B., Partamies, N., and Sangalli, L. (2013). A new automatic method for estimating the peak auroral emission height from all-sky camera images, *Geosci. Instrum. Method. Data Syst.*, 2, 131–144. https://doi.org/10.5194/gi-2-131-2013

Zhang et al. (2019). Wide-field auroral imager onboard the Fengyun satellite, Light: Science & Applications, 8:47. https://doi.org/10.1038/s41377-019-0157-7

**Supplementary material**

# Supplementary Material for "Auroral imaging with combined Suomi 100 nanosatellite and ground-based observations: A case study"

## E. Kallio, Ari-Matti Harri, Olli Knuuttila, Riku Jarvinen, Kirsti Kauristie, Antti Kestilä, Jarmo Kivekäs, Petri Koskimaa, Juha-Matti Lukkari, Noora Partamies, Jouni Rynö, and Mikko Syrjäsuo

**Figure S01**. All-sky camera image at 19:35:10 UT, on January 22nd, 2019, from Svalbard.

**Figure S02**. The magnetometer observations from four stations from Tromsø Geophysical observatory (from top to bottom): Ny-Ålesund (NAL), Longyearbyen (LYR), Hopen Island (HOP), and Bear Island (BJN). The figure shows the horizontal (H) component of the magnetic field in nT. From https://flux.phys.uit.no/geomag.html.

**Figure S03**. The magnetic field's a) X, b) Y and c) Z component 1 minute average data from six IMAGE magnetometer stations: Hornsund (HOR), Hopen Island (HOP), Bear Island (BJN), Nordkapp (NOR), Sørøy (SOR), and Tromsø (TRO) on 2019-01-22 at 16:00-24:00 UT and d) positions of the stations in a map. From https://space.fmi.fi/image/www/.

**Figure S04**. ASC keogram from Sodankylä (67.42°N), Finland, between 2019-01-22 15:00 UT – 2019-01-23 06:00 UT. From https://space.fmi.fi/MIRACLE/ASC/ASC_keograms/SOD.1901/SOD_190122.jpg.

**Figure S05**. Sodankylä Geomagnetic Observatory's riometer observations from six stations: Abisko (30 MHz, 68.40°N 18.90°E), Ivalo (30 MHz, 68.55°N 27.28°E), Sodankylä (30 MHz, 67.42°N 26.39°E), Rovaniemi (32.4 MHz, 66.78°N 25.94°E), Oulu (30 MHz, 65.08°N 25.90°E), and Jyväskylä (32.4 MHz, 62.42°N 25.28°E) on 2019-01-22. From https://www.sgo.fi/Data/Riometer/rioData.php.

**Figure S06**. Seven all-sky images from Kevo (69.76°N), Finland, on 2019-01-22 between 17:59:40 - 18:59:00 UTC. The direction of the magnetic North, East, South and West are top, right, bottom and left in the image, respectively.

**Figure S07**. Three all-sky images from Kilpisjärvi (69.05°N), Finland, on 2019-01-22 between 18:00:39 - 18:59:50 UTC. The direction of the magnetic North, East, South and West are top, right, bottom and left in the image, respectively. Figure 18:59:50 UTC is gamma corrected in order to increase visibility of aurora light.

Figure S01

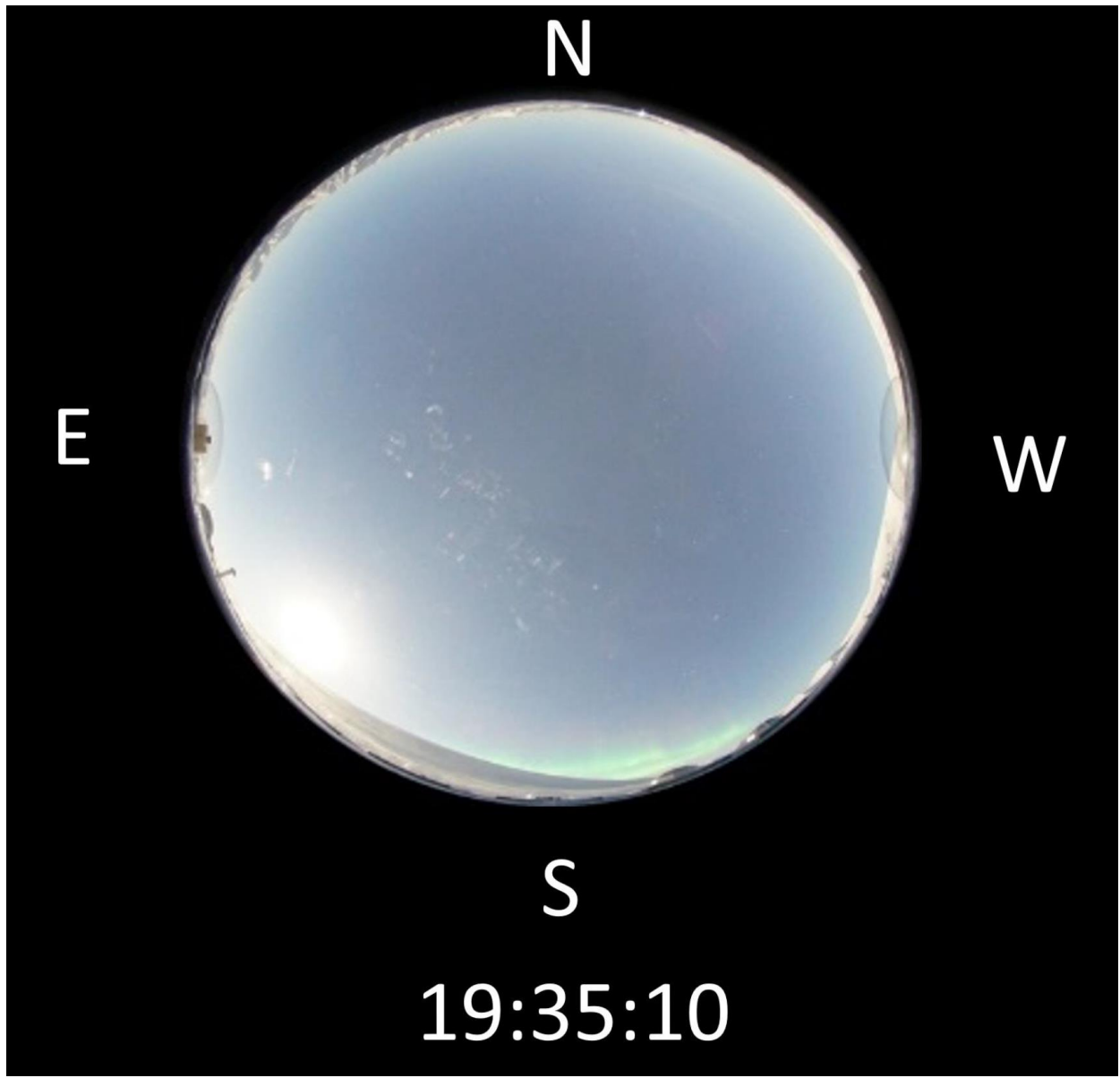

Figure S02

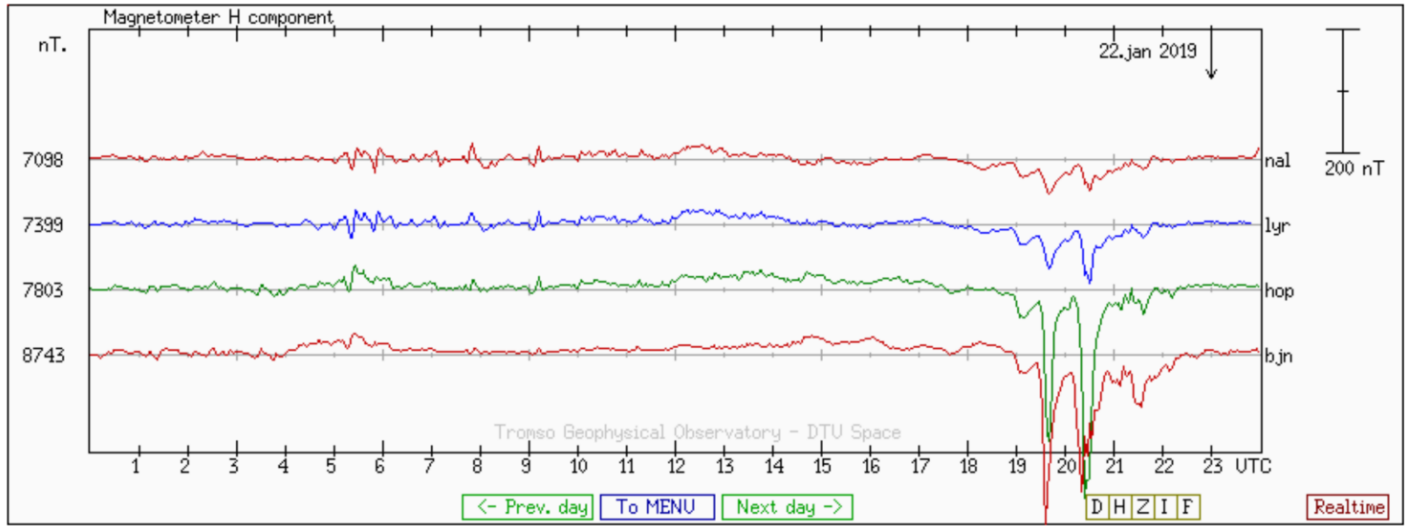

Figure S03

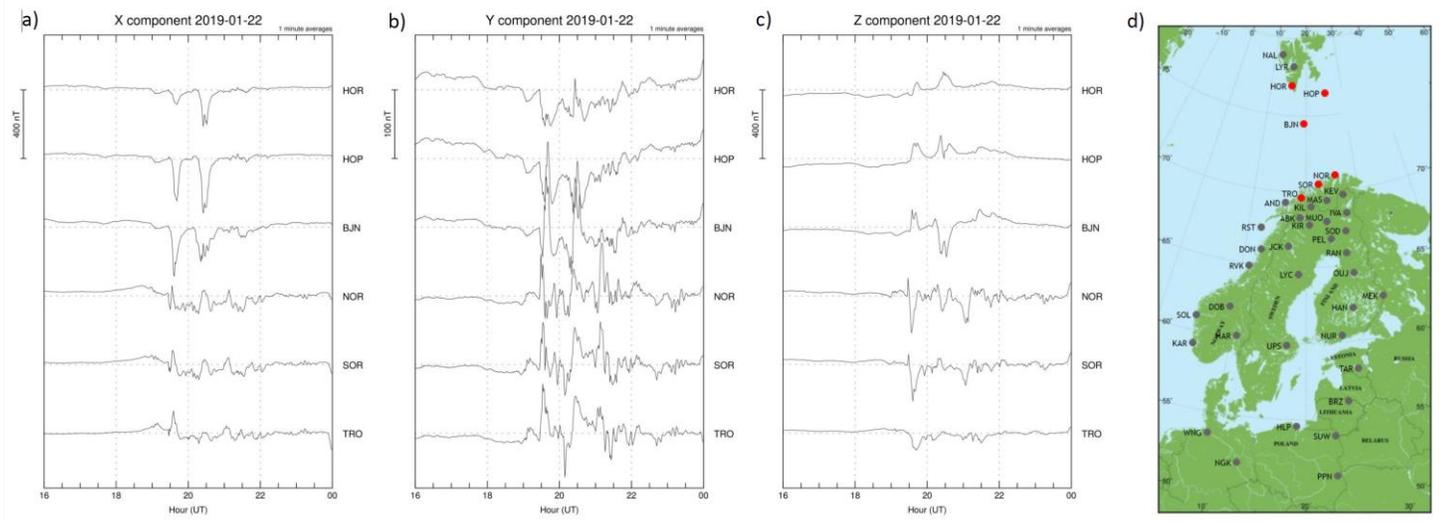

Figure S04

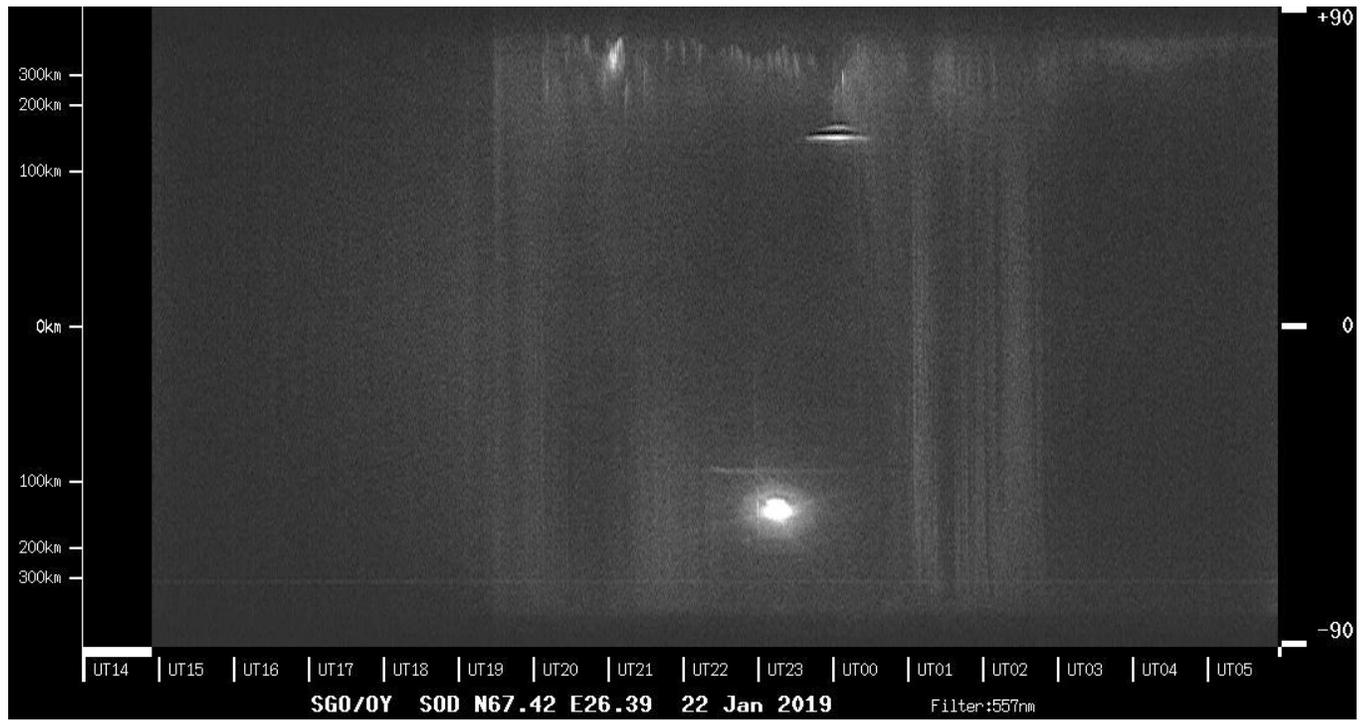

Figure S05

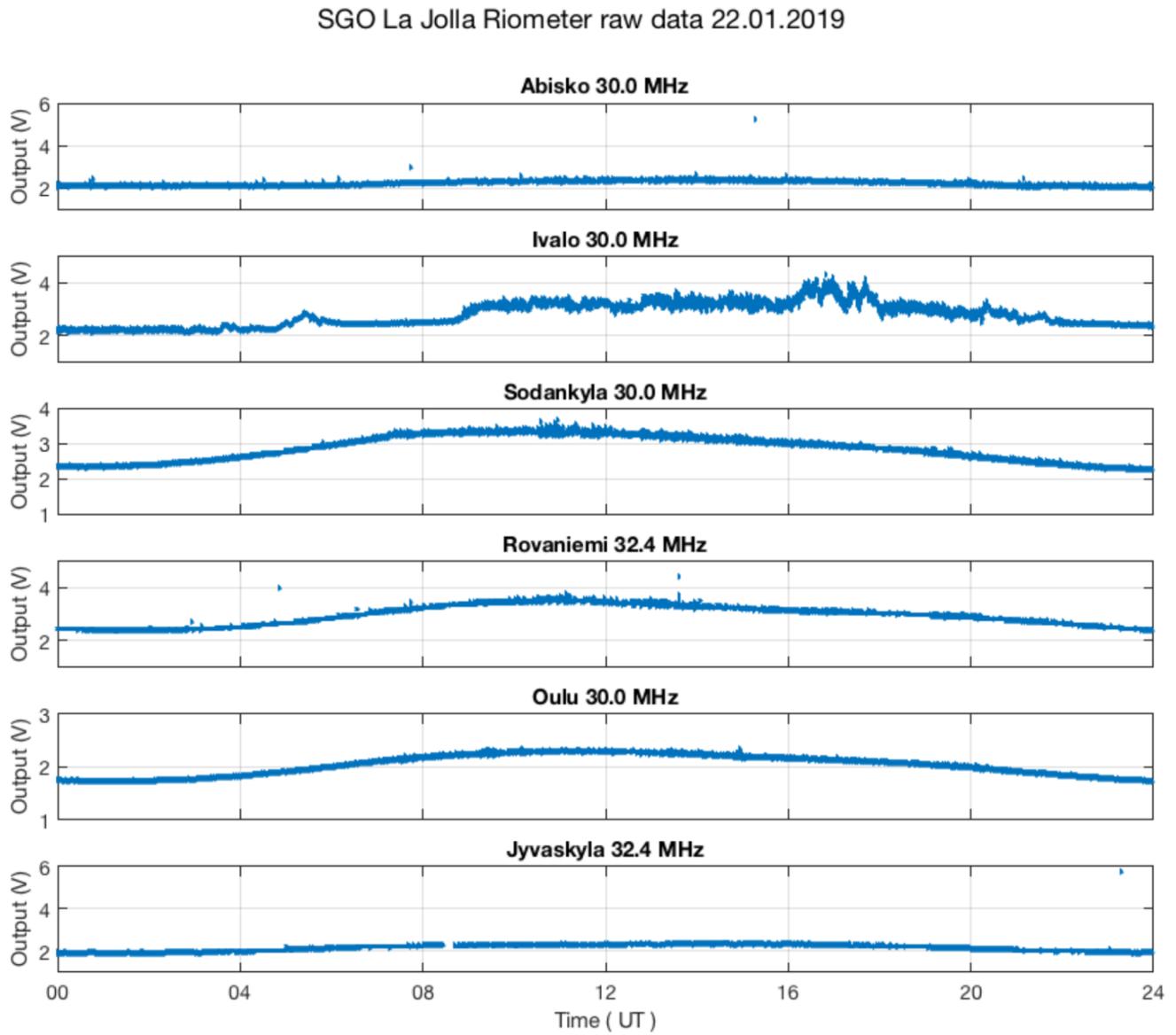

Figure S06

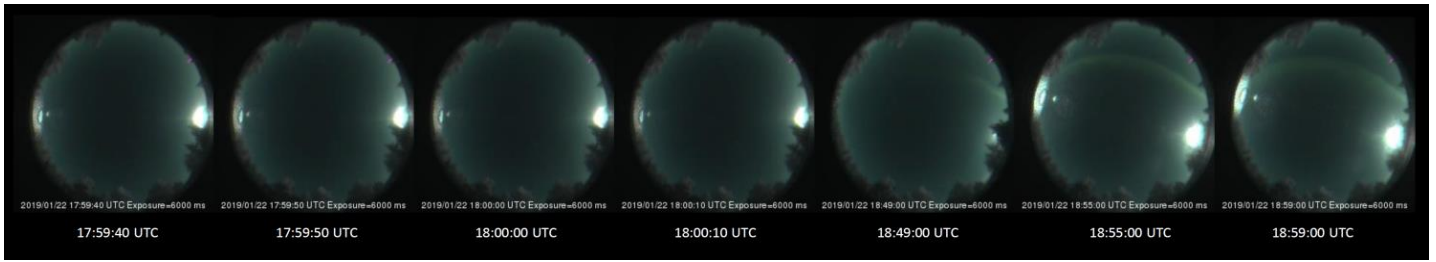

Figure S07

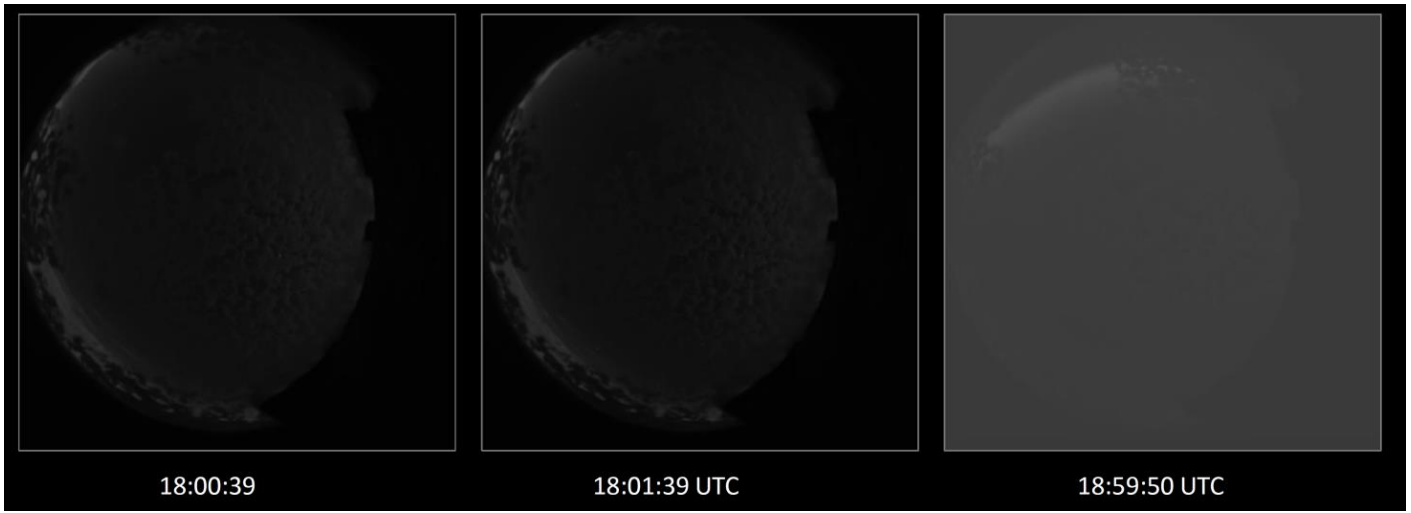